\documentclass[usenatbib]{emulateapj}

\shorttitle{The Evolution of Today's Massive Galaxies}
\shortauthors{A. R. Hill et al.}
\begin{document}
\title{The mass, colour, and structural evolution of today's massive galaxies since $\lowercase{z}\sim5$}

\author{Allison R. Hill$^{1}$, Adam Muzzin$^{2}$, Marijn Franx$^{1}$, Bart Clauwens$^{1}$, Corentin Schreiber$^{1}$, Danilo Marchesini$^{3}$, Mauro Stefanon$^{1}$, Ivo Labbe$^{1}$, Gabriel Brammer$^{4}$, Karina Caputi$^{5}$, Johan Fynbo$^{6}$, Bo Milvang-Jensen$^{6}$, Rosalind E. Skelton$^{7}$, Pieter van Dokkum$^{8}$, Katherine E. Whitaker$^{9,10,11}$}
\affil{$^{1}$Leiden Observatory, Leiden University, P.O. Box 9513, 2300 RA,
Leiden, The Netherlands}
\affil{$^{2}$Department of Physics and Astronomy, York University, 4700 Keele St., Toronto, Ontario, Canada, MJ3 1P3}
\affil{$^{3}$Physics and Astronomy Department, Tufts University, 574 Boston Avenue, Medford, MA, 02155, USA}
\affil{$^{4}$Space Telescope Science Institute, 3700 San Martin Drive, Baltimore, MD 21218, USA}
\affil{$^{5}$Kapteyn Astronomical Institute, University of Groningen, P.O. Box 800, 9700 AV Groningen, The Netherlands}
\affil{$^{6}$Dark Cosmology Centre, Niels Bohr Institute, University of Copenhagen, Juliane Maries Vej 30, DK-2100 Copenhagen, Denmark}
\affil{$^{7}$South African Astronomical Observatory, PO Box 9, Observatory, Cape Town, 7935, South Africa}
\affil{$^{8}$Astronomy Department, Yale University, New Haven, CT 06511, USA}
\affil{$^{9}$Department of Astronomy, University of Massachusetts, Amherst, MA 01003, USA}
\affil{$^{10}$Department of Physics, University of Connecticut, Storrs, CT 06269, USA}
\affil{$^{11}$Hubble Fellow}

\email{hill@strw.leidenuniv.nl}

\begin{abstract}

In this paper, we use stacking analysis to trace the mass-growth, colour evolution, and structural evolution of present-day massive galaxies ($\log(M_{*}/M_{\odot})=11.5$) out to $z=5$. We utilize the exceptional depth and area of the latest UltraVISTA data release, combined with the depth and unparalleled seeing of CANDELS to gather a large, mass-selected sample of galaxies in the NIR (rest-frame optical to UV). Progenitors of present-day massive galaxies are identified via an evolving cumulative number density selection, which accounts for the effects of merging to correct for the systematic biases introduced using a fixed cumulative number density selection, and find progenitors grow in stellar mass by $\approx1.5~\mathrm{dex}$ since $z=5$. Using stacking, we analyze the structural parameters of the progenitors and find that most of the stellar mass content in the central regions was in place by $z\sim2$, and while galaxies continue to assemble mass at all radii, the outskirts experience the largest fractional increase in stellar mass. However, we find evidence of significant stellar mass build up at $r<3~\mathrm{kpc}$ beyond $z>4$ probing an era of significant mass assembly in the interiors of present day massive galaxies. We also compare mass assembly from progenitors in this study to the EAGLE simulation and find qualitatively similar assembly with $z$ at $r<3~\mathrm{kpc}$. We identify $z\sim1.5$ as a distinct epoch in the evolution of massive galaxies where progenitors transitioned from growing in mass and size primarily through in-situ star formation in disks to a period of efficient growth in $r_{e}$ consistent with the minor merger scenario.

\end{abstract}


\section{Introduction}
\label{sec:intro}

The mass growth and structural evolution of today's most massive galaxies is an important tracer of galaxy assembly at early times. These systems are host to the oldest stars, suggesting they were the first galaxies to assemble. Because they are the oldest systems, their progenitors can theoretically be traced to higher redshifts than their low mass counterparts and can be studied from the onset of re-ionization to give a complete history of galactic evolution. Additionally, the most massive systems tend to be the most luminous, and they are the easiest to observe at high redshift with high fidelity. Massive galaxies also provide important constraints on the physics involved in cosmological simulations, as they impose upper limits on growth as well as the efficiency of various feedback mechanisms such as active galactic nuclei, mergers and supernovae. 

Today's massive ($\log{\mathrm{M_{*}/M_{\odot}}}\sim11.5$) galaxies, to first order, are a uniform population. They are homogeneous in morphology and star formation, appearing spheroidal and have low specific star formation rates, and high quiescent fractions \citep[e.g.,][]{thomas2005, gallazzi2005, kuntschner2010, thomas2010, cappellari2011, mortlock2013, moustakas2013, ilbert2013, muzzin2013b, davis2014, mcdermid2015}. In contrast to today's massive galaxies, massive galaxies at high redshift show increasing diversity \citep[e.g.,][]{franx2008, vandokkum2011}. With increasing redshift, massive galaxies become increasingly star forming \citep[e.g.,][]{papovich2006, kriek2008, vandokkum2010, brammer2011, bruce2012, ilbert2013, muzzin2013b, patel2013a, stefanon2013, barro2014, duncan2014, marchesini2014, toft2014, vandokkum2015, barro2016, man2016, tomczak2016}, and the massive galaxies which are identified as quiescent at high redshift are structurally distinct from their low redshift counterparts as seen in their small effective radii ($r_{e}$) and more centrally concentrated stellar-mass density profiles \citep{daddi2005, trujillo2006, toft2007, cimatti2008, vandokkum2008, damjanov2009, newman2010, szomoru2010, williams2010, vandesande2011, bruce2012, muzzin2012, oser2012, szomoru2012, szomoru2013, mclure2013, vandesande2013, newman2015, straatman2015, hill2016}. 

Although the central regions of massive galaxies contain a higher fraction of the total mass at high redshift,  their central stellar densities show remarkably little evolution between $z\approx2-3$ and $z=0$ \citep[e.g.,][]{bezanson2009, vandokkum2010, toft2012, vandesande2013, patel2013a, belli2014a, vandokkum2014, williams2014, whitaker2016} with the majority of stellar-mass build-up occurring in the outer regions (with galaxies growing in an `inside-out' fashion). This mass assembly is thought to occur via minor, dissipation-less mergers; a scenario which is able to account for the size growth, while leaving the interior regions relatively undisturbed \citep[e.g.,][]{bezanson2009, naab2009, hopkins2010, trujillo2011, newman2012, hilz2013, mclure2013}. The aims of the present study are to determine whether these trends continue to high redshifts and to identify the epoch when galaxies' central regions assemble their mass. 

\begin{figure*}[t]
  \begin{center}
  \includegraphics[width=\textwidth]{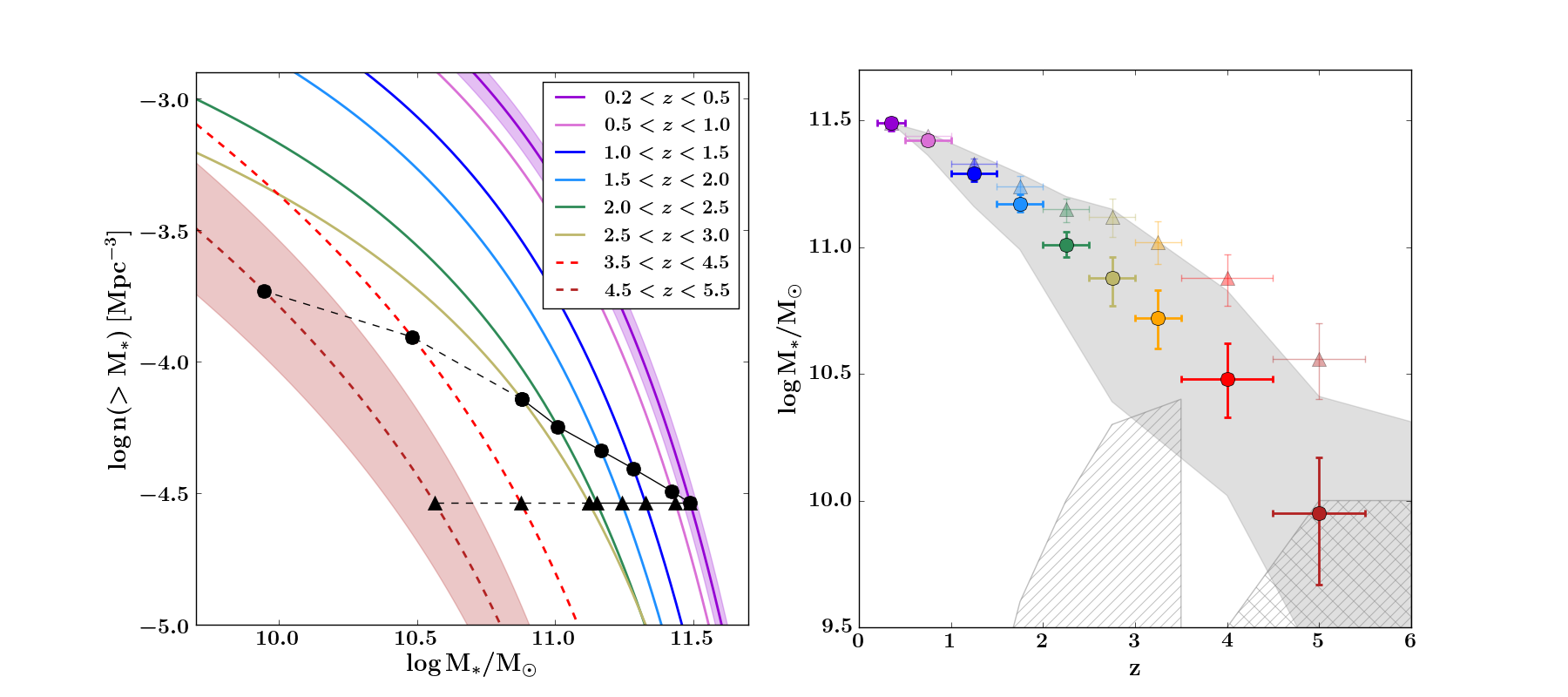}
  \caption{\textit{Left}: Integrated mass functions as a function of stellar mass for different $z$ ranges. Solid and dashed lines indicate the mass functions of \citet{muzzin2013b} and \citet{grazian2015}, respectively, with colour illustrating the redshift. Uncertainties in the mass functions resulting from uncertainties in the photo-$z$'s, SFH and cosmic variance are shown for the highest- and lowest-$z$ (for clarity). Black circles indicate the cumulative number density selection of \citet{behroozi2013}, with black triangles showing a fixed-cumulative number density selection for comparative purposes. \textit{Right:} The mass evolution of the progenitors of a $\log(M/M_{\odot})=11.5$ galaxy at z=0.35. As in the left panel, the circles and triangles show an evolving and fixed cumulative number density selection. The difference between the circles and the triangles illustrate the bias, especially at $z>2$, resulting from a fixed number density selection. The error-bars in the y-axis are the uncertainties resulting from the mass function. The error-bars in the x-axis represent the redshift range considered. The solid grey-regions indicate the $1-\sigma$ range from \citet{behroozi2013}, and the hatched regions represent our estimated mass completeness limits which are discussed in Sec.~\ref{sec:data}.}
  \label{fig:progenitors}
  \end{center}
\end{figure*}

Obtaining a census of massive galaxies across a broad redshift range is technically challenging, as they have low number densities on the sky \citep{cole2001, bell2003, conselice2005, marchesini2009, bezanson2011, caputi2011, baldry2012, ilbert2013, muzzin2013b, duncan2014, tomczak2014, caputi2015, stefanon2015, huertas2016} and their rest-frame optical emission shifts into the near-infrared (NIR) at intermediate redshifts. To study the evolution of massive galaxies across cosmic time, as a population, necessitates deep and wide NIR surveys to both probe large volumes and obtain rest-frame optical emission to significant signal-to-noise (S/N).  

In this study we use stacking analysis to obtain high-fidelity profiles of the progenitors of massive galaxies out to significant radii (at low $z$, $r>60~\mathrm{kpc}$). We take advantage of the unparalleled combination of depth and area in the third data release of the UltraVISTA survey \citep{mccracken2012} to study the structural evolution of massive galaxies out to $z=3.5$. Due to incompleteness in UltraVISTA at the highest redshifts considered in this study, we also use the deeper CANDELS F160W data from the 3DHST photometric catalogs \citep{brammer2012a, skelton2014, momcheva2016} to extend the redshift coverage to $z=5$. This is a significant gain in redshift over previous studies, and provides the most extensive redshift range over which the profiles of massive galaxies have been traced.

\section{Sample Selection}
\label{sec:data}

\subsection{Number-density selection}
\label{sec:selection}

Linking the progenitors of present day galaxies to their high redshift counterparts is challenging, as the merger and star formation history (SFH) of any individual galaxy is not well constrained. One way to circumvent these issues is to assume that galaxies maintain rank-order across cosmic time (i.e., the most massive galaxies today will have been the most massive galaxies yesterday, cosmologically speaking). This assumption predicts a constant co-moving number-density with redshift, an outcome used by \citet{vandokkum2010} to trace the mass and size growth of galaxies from $z=2$ (corresponding to $n=2\times10^{-4}~\mathrm{Mpc^{-3}dex^{-1}}$). Subsequent studies have used the same assumptions to select progenitors based on a constant \textit{cumulative} number density \citep[e.g.,][]{bezanson2011, brammer2011, papovich2011, fumagalli2012, vandokkum2013, patel2013a, ownsworth2014, morishita2015}, which has the advantage over its non-cumulative counterpart of being single valued in mass. 

The selection of progenitors and their descendants at a constant cumulative number density implicitly assumes that mergers and in-situ star-formation do not broadly effect rank-order, an assumption which has been shown to result in systematically biased progenitor selection \citep{behroozi2013, leja2013, torrey2015}. To account for the affects of mergers on the progenitor mass, we utilize an evolving cumulative number density selection following the prescription of \citet{behroozi2013}, who use halo-abundance matching within a $\Lambda\mathrm{CDM}$ cosmology to connect progenitors and their descendants. It is important to note, that we have used the prescription to trace \textit{progenitors} of low redshift massive galaxies, not the \textit{descendants} of high redshift massive galaxies, of which the former yields a steeper evolution in cumulative number density due to the shape of the halo mass function, and scatter in mass accretion histories \citep[see][]{behroozi2013, leja2013}.

\subsection{The implied stellar mass growth of the progenitors of massive galaxies since $z\sim5$}
\label{sec:mass_growth}

In Fig.~\ref{fig:progenitors} we show the integrated Schecter fits of the mass functions of \citet{muzzin2013b} between $0.2<z<3.0$, and \citet{grazian2015} between $3.5<z<5.5$. These mass functions are based on photometric redshifts determined via ground and space based NIR imaging from the UltraVISTA and CANDELS surveys respectively. In the left-panel of Fig.~\ref{fig:progenitors}, we show our evolving cumulative number density selection based on the abundance matching of \citet{behroozi2013}.  The masses implied from a fixed-cumulative number density selection are also shown to illustrate the effect of the bias when the effects of mergers are ignored in the selection. In the right-panel of Fig.~\ref{fig:progenitors} the implied progenitor masses from the left-panel are plotted for both the fixed and evolving cumulative number density selection, as a function of redshift. The error bars are the uncertainties from the mass functions, which take into account the uncertainties in the photometric redshifts, SFHs, and cosmic variance.  The solid grey region represents the scatter in the number densities from the abundance matching of \citet{behroozi2013}, and the hatched regions illustrate an estimate of the mass completeness which is discussed in detail in Sec.~\ref{sec:data}. 

Below $z=2$, Fig.~\ref{fig:progenitors} shows that both constant and evolving cumulative number density selections yield progenitor masses which are consistent within the uncertainties in the mass-functions. However beyond $z=2$, the bias in the fixed cumulative number density becomes significant, and over-predicts the median progenitor mass. Using the abundance matching technique, we see an overall increase in stellar mass of $1.5~\mathrm{dex}$ since $z\sim5$. Our fractional mass growth out to $z=3$ is consistent within the uncertainties with \citet{marchesini2014} who use the same abundance matching selection for ultra-massive $\log(M_{*}/M_{\odot})\sim11.8$) descendants, and with \citet{ownsworth2014}, who use a constant cumulative number density selection which is corrected for major mergers to trace progenitors. Using their correction, they find $75\pm9\%$ of the descendant mass is assembled after $z=3$, which is consistent with $\sim80\%$ which we find in the current study. 

We note that in Fig.~\ref{fig:progenitors} we have selected a progenitor mass for a redshift bin between $3.0<z<3.5$ (orange point), even though we have indicated no mass-function for this redshift. The mass-function from \citet{muzzin2013b} for this redshift range proved to be unreliable for the mass considered due to incompleteness from UltraVISTA DR1 (the source catalog used in generating the mass functions). However, with the deeper exposures from the third data release (DR3) of UltraVISTA, we are complete to the progenitor masses considered out to $z=3.5$. To calculate the expected progenitor mass between $3.0<z<3.5$, we linearly interpolated the mass between adjacent redshift bins. We also observe a trend of the uncertainties in the mass function monotonically increasing from low to high redshift. Thus, we similarly linearly interpolated the uncertainties to estimate the uncertainty in mass for $3.0<z<3.5$ due to uncertainties in photo-$z$, SFH and cosmic variance. We also use the uncertainties in the progenitor mass selection as the upper and lower mass bounds for the galaxies that contribute to the resulting stack, thus we select a larger range of masses at higher redshift, than at lower redshift. 

It has been shown that the \citet{behroozi2013} prescription for selecting progenitors performs well in terms of recovering the \textit{average} stellar mass of the progenitors of present-day high-mass galaxies, however this method fails in capturing the diversity in mass of all progenitors as implied by simulations \citep[e.g.,][]{torrey2015, clauwens2016, wellons2016b}, which also predict that the scatter in progenitor masses tends to increase with redshift. Given this large scatter, there is no guarantee that the evolution of other galaxy properties, such as size, will follow from the \citet{behroozi2013} selection. However, in an upcoming paper (Clauwens et al., in prep) we will show that for the property of interest in our study (i.e. the average radial build-up of stellar mass for the progenitors of massive galaxies), the \citet{behroozi2013} selection yields average agreement with progenitors within the EAGLE simulation. 

\subsection{Data}
\label{sec:data}

\subsubsection{UltraVISTA}

In order to study the evolution of the average properties of massive galaxies, it was necessary to utilize both wide field ground-based, and deep space-based imaging for our stacking analysis. Massive galaxies ($\log(M_{*}/M_{\odot})\sim11$) are exceedingly rare objects, with low number densities ($\sim10^{-5}~\mathrm{Mpc^{-3}}$) on the sky \citep[e.g.,][]{cole2001, bell2003, baldry2012, muzzin2013b, ilbert2013, tomczak2014, caputi2015, stefanon2015}, and require wide-field surveys to characterize a significant population. To that end, we utilize the NIR imaging from the DR3 of the UltraVISTA survey \citep{mccracken2012} for our stacking analysis.

The DR3 UltraVISTA catalog (Muzzin et al., in prep) is a K-selected, multi-band catalog constructed from the UltraVISTA survey. Briefly, the survey covers the COSMOS field with a total area of $1.7~\mathrm{deg^{2}}$, with deep imaging in the $Y, J, H$ and $Ks$ bands. The survey also contains ultra-deep stripes with longer exposures which cover a $0.75~\mathrm{deg^{2}}$ area, and also includes imaging in the VISTA NB118 NIR filter \citep{milvang2013}. The newest data release is constructed with the same techniques as the DR1 30-band catalog \citep{muzzin2013a}, with the inclusion of new and higher-quality data to determine photo-z's, and stellar population parameters. The DR3 survey depths in the ultra-deep stripes are $\sim1.4$ magnitudes deeper than DR1 (with $5\sigma$ limiting magnitudes in the ultra-deep regions of 25.7, 25.4, 25.1, and 24.9 in $Y, J, H$ and $Ks$). 

Several other datasets have also been added since the first data release including 5 CFHTLS filters, $u^{*}g^{\prime}r^{\prime}i^{\prime}z^\prime$, as well as 2 new Subaru narrow bands (NB711, NB816). Most importantly for this analysis we also include the latest data from SPLASH \citep{capak2012} and SMUVS (PI Caputi; Ashby et al., in prep). These are post-cryo \textit{Spitzer}-IRAC observations that improve the $\mathrm{[3.6]}$ and $\mathrm{[4.5]}$ depth from 23.9 to 25.3. Overall this is a $38$-band catalog (compared to 30 in \citealt{muzzin2013a}), and the substantial increase in depth in the $Y,J,H,Ks$, $\mathrm{[3.6]}$ and $\mathrm{[4.5]}$ bands make it a powerful dataset for studying massive galaxies at intermediate and high redshifts.
 
In the right panel of Fig.~\ref{fig:progenitors} we have indicated our estimated mass completeness limits with the filled hatched regions. To estimate our mass completeness at $z<4$, we used the limits on the mass functions from \citet{muzzin2013b} (which were derived using UltraVISTA DR1), and adjusted the mass limit according to the gain in K-band depth (the K-band limit is 1.5 magnitudes deeper between DR1 and DR3) assuming a constant mass-to-light ratio.  Since galaxy mass-to-light ratios decrease with redshift \citep[e.g.][]{vandesande2015}, this likely represents a conservative estimate of the limiting mass at high redshifts. 

\subsubsection{CANDELS}

As UltraVISTA DR3 is only mass complete for our selection out to $z=3.5$, we use the reddest band available from CANDELS in order to explore redshifts unobtainable through UltraVISTA. We select galaxies using the photometric data products from the 3DHST survey \citep{brammer2012a, skelton2014} from all 5 CANDELS fields. As an estimate of our mass completeness in CANDELS, we adopt the limiting mass derived from the $75\%$ magnitude completeness limit ($F160W=25.9$) in the shallower pointings in the GOODS-S and UDS fields as described in \citet{grazian2015}. They estimated their mass completeness using the technique of \citet{fontana2004}, which assumes the distribution of mass-to-light ratios immediately above the magnitude limit holds at slightly lower fluxes, and compute the fraction of objects lost due to large mass-to-light ratios. The estimated completeness for CANDELS is indicated in the right panel of Fig.~\ref{fig:progenitors} as the grey cross-hatched region.

Although the aforementioned estimates of mass completeness take into account galaxies with varied mass-to-light ratios, it is worth stressing inherent uncertainties when determining mass limits at high redshift. At $z>3.5$, we increasingly rely on photometric redshifts, as high-fidelity spectroscopic redshifts are fewer in number \citep{grazian2015}. In addition, sub-mm galaxies (SMGs) likely account for at least a fraction of the progenitors of massive galaxies at high redshift \citep[e.g.,][]{toft2014}, and they have been shown to have high optical extinction \citep[e.g.,][]{swinbank2010,couto2016}. As the progenitors selected at $z>3.5$ of this study tend to be less massive than a typical SMG, we do not expect that they will form a significant fraction of the sample. However, we cannot rule out a tail of less, but still obscured sources to lower masses in the distribution of SMGs. This would have the effect of biasing our high redshift progenitor selection to bluer, less-obscured sources.


\begin{deluxetable}{lcc}
\tabletypesize{\scriptsize}
\tablecaption{Number of galaxies in each redshift range by catalog}
\tablewidth{0pt}
\tablehead{
\colhead{$z$-range} & \colhead{$UVISTA$} & \colhead{$3DHST$} 
}
\startdata
$0.2<z<0.5$ & 16 & 0 \\
$0.5<z<1.0$ & 56  & 5 \\
$1.0<z<1.5$ & 96 & 22 \\
$1.5<z<2.0$ & 166 & 31 \\
$2.0<z<2.5$ & 276 & 79 \\
$2.5<z<3.0$ & 466 & 104 \\
$3.0<z<3.5$ & 160 & 69 \\
$3.5<z<4.5$ & ... & 110 \\
$4.5<z<5.5$ & ... & 154\tablenotemark{*}
\\
\enddata
\tablenotetext{*}{We are incomplete in mass for this point}
\tablecomments{Above are the number of galaxies found within the mass ranges outlined in Fig.~\ref{fig:progenitors}.}
\label{tab:number}
\end{deluxetable}


Table~\ref{tab:number} provides a summary of the number of galaxies in the given redshift range, at the implied mass as determined from our evolving cumulative number density selection (see Sec.~\ref{sec:selection}) from both the UltraVISTA and 3DHST catalogs. In order to boost the number of galaxies in UltraVISTA, we have used galaxies from both the deep (DR1) and ultra-deep (DR3) catalog out to $2.0<z<2.5$ where we are complete in mass for the shallower catalog (DR1). For the $3.0<z<3.5$ bin, we have only utilized the DR3 catalog, as we are incomplete in DR1. As evident from Table~\ref{tab:number}, UltraVISTA has a larger population of massive galaxies at low redshift, while there are 0 galaxies in all 5 CANDELS fields which are massive ($\log(M_{*}/M_{\odot})\sim11.5$) at $z=0.35$, and only 5 galaxies in the next highest redshift bin. However CANDELS is crucial to continue the progenitor selection beyond $z>3.5$ as we are mass incomplete in this region with UltraVISTA. Additionally, as galaxies had smaller $r_{e}$ at high redshift (see discussion in Sec.~\ref{sec:intro} and references therein), the space-based seeing of CANDELS is necessary to properly map the density profiles at these epochs. Thus we utilize both data sets in our analysis. 

\section{Rest-Frame Color Evolution}

Cumulative number density selection is a method which selects solely on stellar mass, and is therefore blind to other galaxy properties such as levels of star-formation activity. A simple, but effective way to establish star-forming activity in a population of galaxies is to observe where they are located in rest-frame $U-V$ and $V-J$ color space, commonly referred to as a $UVJ$-diagram. First proposed by \citet{labbe2005}, it is observed that galaxies exhibit a bi-modality in rest-frame $UVJ$ colour space which is correlated with the level of obscured and unobscured star formation. Actively star-forming and quiescent galaxies separate into a `blue' and `red' sequence in the $UVJ$-diagram \citep[e.g.,][]{williams2009, williams2010, whitaker2011, fumagalli2014, yano2016}. 

\begin{figure*}
  \begin{center}
  \includegraphics[width=\linewidth]{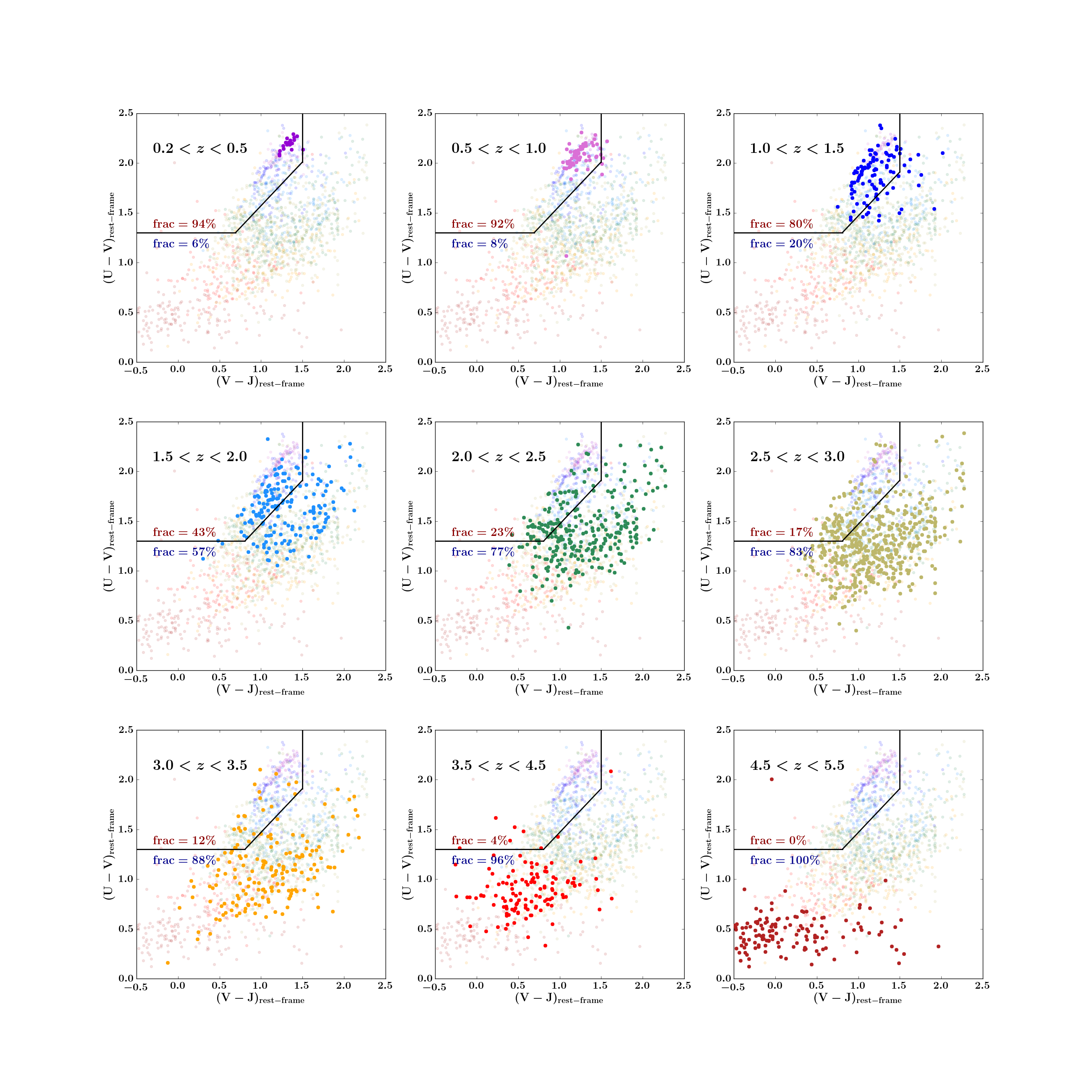}
  \caption{Rest-frame $UVJ$ diagrams separated according to redshift bin, for all galaxies used in the stacked images. The redshift increases from top-left to bottom-right. Each panel highlights the galaxies which are both in the redshift, and mass ranges considered in Fig.~\ref{fig:progenitors}, as well as the full sample re-plotted, but washed out to illustrate how each bin relates to the over-all sample. The star-forming/quiescent division from \citep{muzzin2013b} in $UVJ$ colour is over-plotted in black. The first seven panels contain galaxies drawn from the UltraVISTA DR3 catalog, and the 8th and 9th panels are from CANDELS-3DHST. There is a clear progression in colour evolution from one redshift bin to the other as galaxies start out in the lower-left region of the diagram, and progress along the star-forming sequence before ending at the tip of the red-sequence. It is important to note that in the highest-$z$ panel we are incomplete in mass and are likely biased towards bluer galaxies.}
  \label{fig:uvj}
  \end{center}
\end{figure*}

In Fig.~\ref{fig:uvj}, we plot the rest-frame $U-V$ and $V-J$ colours for all redshift bins to provide a diagnostic of star-formation activity within each stack. Each of the nine panels represents a different redshift range, with galaxy masses selected according to their expected evolving cumulative number density (see Fig.~\ref{fig:progenitors}). The first seven panels are galaxies from UltraVISTA DR3, and the last two panels contain galaxies from the 3DHST photometric catalog. It is important to note that we are mass incomplete for the $4.5<z<5.5$ bin (see Fig.~\ref{fig:progenitors}). However we have chosen to include it as part of our analysis, with the caveat that we are likely biased towards bluer galaxies. Overlaid in each panel are the colour selections used by \citet{muzzin2013b} to separate quiescent and star forming sequences. 

As one progresses in redshift, it becomes apparent from Fig.~\ref{fig:uvj} that the number of galaxies selected dramatically increases. This is a result of two competing effects. The first, is that the size of our mass range becomes progressively larger with redshift, as seen in the error bars on the right-panel of Fig.~\ref{fig:progenitors}. By selecting in a wider mass range, we will inevitably select more galaxies. The second effect is that as the number densities of progenitors increases with redshift, we are progressing towards the lower mass end of the mass-functions \citep{ilbert2013, muzzin2013b, grazian2015}. Thirdly, at low redshift, the probed co-moving volume is also smaller than at high redshift. The combined effect is to have our lowest redshift, and least populated stack contain only 16 galaxies, whereas our most populated stack at $2.5<z<3.0$ contains 276 objects (Table~\ref{tab:number}).

\begin{figure}
  \begin{center}
  \includegraphics[width=\linewidth]{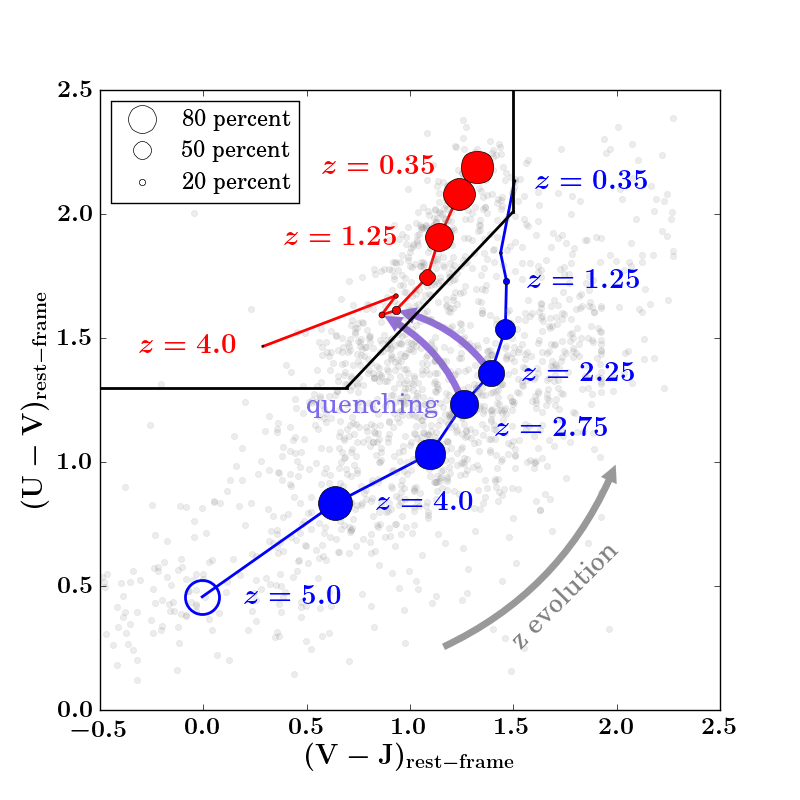}
  \caption{Above is the average rest-frame $UVJ$ colour evolution for the progenitors of the quiescent (red symbols) and star-forming (blue symbols) progenitors. The entire sample is plotted in small grey symbols to best illustrate the scatter. The size of the red and blue symbols indicates the quiescent/star-forming fraction (e.g., a large red circle correspond to a high quiescent fraction, and a small blue circle corresponds to a low star-forming fraction). The redshift evolution proceeds from bottom-left to top-right. Purple arrows indicate the direction of quiescence and are labelled for points which bracket a quiescent fraction of $20\%$. The $z=5$ point is plotted as an open circle to remind the reader that we are incomplete in that redshift bin, and are likely biased to bluer galaxies.}
  \label{fig:uvj_average}
  \end{center}
\end{figure}

The most prominent trend in Fig.~\ref{fig:uvj} comes in the colour evolution of the progenitors across redshift. They begin very blue in both $U-V$ and $V-J$ in the lower-left of the star-forming sequence and progress red-ward along the star-forming sequence to the upper-right until $2.5<z<3.0$ before reddening in $V-J$ and joining the quiescent sequence. Assuming our number density selection is valid, this represents a true evolution in $UVJ$ colour. 

Fig.~\ref{fig:uvj_average} show the average $UVJ$ colour evolution for each redshift bin, separated into star-forming and quiescent progenitors, and highlights explicitly the trends observed in Fig.~\ref{fig:uvj}. In this figure, we see most of the early ($z>3$) colour evolution is driven by the star-forming progenitors. At $z<3$, star-forming progenitors are beginning to quench in large numbers and the two tracks are broadly parallel until $z<1$ where the quiescent progenitor fractions are high, and $UVJ$ colour evolution is driven by the quiescent progenitors. This seems to indicate that massive galaxies begin their existence as star forming galaxies, which progress along the blue sequence (via aging of the stellar populations, and increase in stellar mass through star formation), before quenching and joining the red sequence. 

The progression in the $UVJ$-diagram between $0.2<z<3.0$ is qualitatively similar to \citet{marchesini2014} who tracked the progenitors of local ultra-massive ($\log(M_{*}/M_{\odot})\sim11.8$) galaxies, with the main difference being that this study contain galaxies which are bluer than those of \citet{marchesini2014}. The origin of this difference is rooted in the fact that we select progenitors for a lower local mass galaxy ($\log(M_{*}/M_{\odot})\sim11.5$). Our galaxies in the higher-$z$ bins are also bluer than the sample of \citet{ownsworth2016}, who select progenitors of massive galaxies based on fixed-cumulative number density. As previously discussed, a fixed cumulative number density selection will yield progenitors which are systematically more massive and thus, redder in $U-V$ and $V-J$ colours, and the inconsistencies in galaxy properties between the samples is likely attributed to differences in stellar mass. 

The progression of galaxies between different redshift bins within Fig.~\ref{fig:uvj} and Fig.~\ref{fig:uvj_average} already provides clues as to the structure of the galaxies within them. Numerous studies find that galaxies in the quiescent region of the $UVJ$-diagram tend to have higher $n$ and smaller $r_{e}$ \citep[e.g.,][]{williams2010, patel2012, yano2016}. However, those analyses were for galaxies at fixed masses and did not connect progenitor to descendent, and therefore do not make a direct evolutionary link. In the next section we examine the size and structural evolution of the galaxies selected using the cumulative number density method.

\section{Evolution in Far-Infrared Star Formation Rates}

In Fig.~\ref{fig:uvj} and Fig.~\ref{fig:uvj_average}, we see evidence that the evolution of massive galaxies can be broadly separated into two epochs. At $z>1.5$, galaxies have colours which are consistent with growth mainly through in-situ star formation. At $z<1.5$, galaxy colours are consistent with quenched systems, with mergers becoming the dominant mechanism for growth. We can estimate this epoch more directly by comparing star formation rates to the mass assembly implied from the evolving cumulative number density selection. 

\begin{figure}
  \begin{center}
  \includegraphics[width=\linewidth]{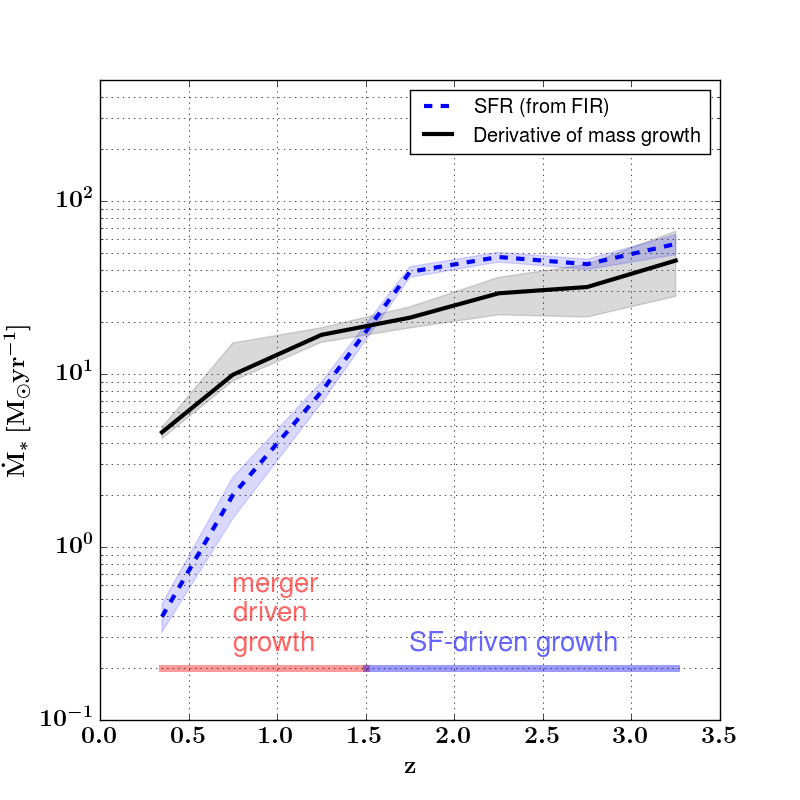}
  \caption{The \textit{FIR} implied star formation rates (dashed blue line), compared to the derivative of the mass-redshift evolution (solid black line), with their associated uncertainties (shaded regions). The implied mass assembly from star formation is higher than the derivative of the mass evolution, at $z>1.5$, and lower at $z<1.5$. At low redshifts, we see the star formation rates drop precipitously, and that mass assembly cannot be proceeding via in-situ star formation, and growth is likely merger driven.}
  \label{fig:sfr}
  \end{center}
\end{figure}


Fig.~\ref{fig:sfr} shows the SFR plotted against the derivative of the progenitor mass growth from the right panel of Fig.~\ref{fig:progenitors}. The SFRs are calculated from far-infrared (FIR) luminosities, which are derived from stacks which include \textit{Spitzer} $24~\micron$, and \textit{Herschel} PACS and SPIRE bands. For each UltraVISTA stack, FIR stacks were generated in the same manner as described in \citet{schreiber2015}. From Fig. 4, of \citet{schreiber2015}, we see that we do not have sufficient numbers of galaxies at $z>3.5$ with the CANDELS data to expect a FIR detection. Thus we only calculate SFRs out to $z=3.5$. The FIR luminosities were converted to SFRs via the relation from \citet{kennicutt1998}, with a factor of 1.6 correction to convert between the Salpeter IMF used in \citet{kennicutt1998}, to the Chabrier IMF used for the DR3 catalog. 

In order to more directly compare the net stellar mass growth as implied from the abundance matching technique to the stellar mass growth from star formation, a $50\%$ conversion factor has been applied to the SFR to account for stellar mass which is lost in outflows from stellar winds \citep[see][]{vandokkum2008, vandokkum2010}. From Fig.~\ref{fig:sfr}, we see that SF is able to account for all of the stellar mass growth at $z>1.5$, with little to no contribution from mergers. In contrast, the SFR at $z<1.5$ are insufficient to explain the mass growth, suggesting stellar mass is accreted via mergers. 

Between $1.5<z<2.5$, the stellar mass growth predicted from star formation is greater than what is found from the abundance matching techniques by $0.1-0.2~dex$. This discrepancy is also seen in model and observation comparisons (see \citealt{somerville2015, madau2014}), with potential for the FIR SFRs to be over estimated during this epoch (see \citealt{madau2014} and discussion therein). In spite of this, the FIR SFR support the notion that massive galaxies grow via star-formation until $z\sim1.5$, where merger driven growth dominates, consistent with the rest-frame $UVJ$ colours, and what is found in the literature (see Sec.~\ref{sec:intro} and references therein). 

\section{Analysis}
\label{sec:analysis}

\subsection{Stacked Images}
\label{sec:stacks}


\begin{figure*}
  \begin{center}
  \includegraphics[width=\linewidth]{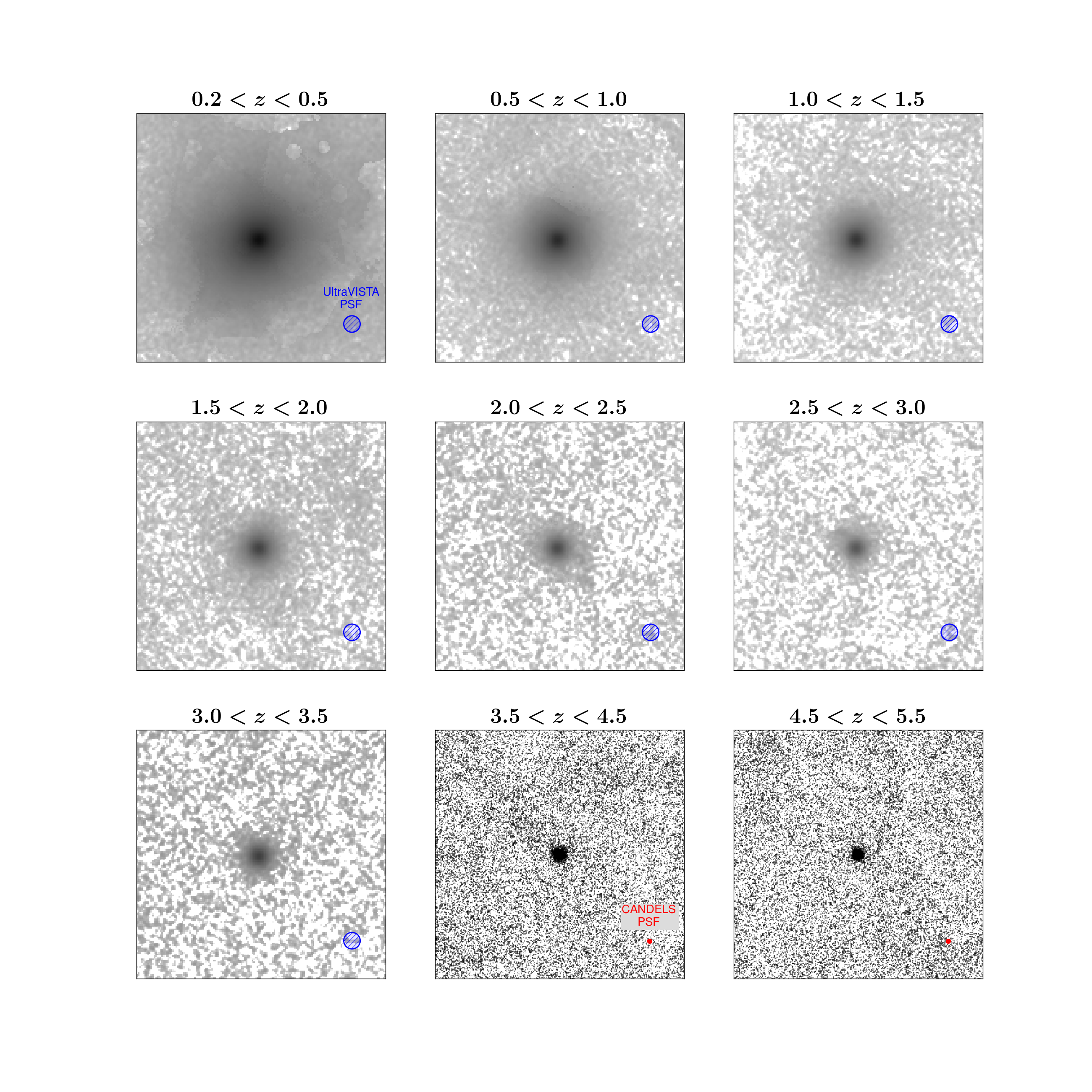}
  \caption{Sample stacked images for each redshift bin. The first seven panels contain stacks from the UltraVISTA data, with each panel containing a stack from the band which is closest to the rest-frame $0.5\micron$. The $Y$-band is chosen for stacks at $z<1$, the $J$-band at stacks $1<z<2$, the H-band at $2<z<3$, the $Ks$-band at $3.0<z<3.5$. The UltraVISTA stacks are all displayed at the same colour scale to high-light differences in background and S/N. The last two panels are $F160W$ stacks, and are plotted at the same scale to each other, although different from the UltraVISTA images for clarity, as the background is much higher in the higher-$z$ bins. Overlaid on each panel is a circle which represents the size of the PSF of the data which contributed to the stack, with the ground-based data having a significantly larger PSF than the HST data.}
  \label{fig:stacks}
  \end{center}
\end{figure*}


For galaxies at $z<3.5$, images were stacked using $48^{\prime\prime}\times48^{\prime\prime}$ cutouts taken from the UltraVISTA mosaics, which contain both deep and ultra-deep stripes. For each cut-out, SEDs were generated using the ancillary data available in the UltraVISTA and CANDELS source catalogs. These SEDs were used to flag potential active galactic nuclei (AGN), which were removed from the resultant stack. The individual cut-outs were also visually inspected to remove objects which were identified as doubles, or triples (i.e., were not separated by SExtractor; \citealt{bertin1996}) or in close proximity to saturated stars to maintain image fidelity. In total, $<4\%$ of the entire sample was discarded. 

Cutouts were centered using coordinates taken from the UltraVISTA DR1 (only deep stripes) and DR3 (only ultra-deep stripes) catalogs, with cubic spline interpolation performed for sub-pixel shifting. For galaxies at $z>3.5$, images were stacked using $24^{\prime\prime}\times24^{\prime\prime}$ cutouts, taken from the 5 CANDELS fields (AEGIS, COSMOS, GOODS-S, GOODS-N and the UDS), with images centred using the coordinates from the 3DHST photometric catalogs \citep{brammer2012a, skelton2014} with sub-pixel shifting also performed using cubic spline interpolation.

From these cutouts, bad-pixel masks were also constructed using SExtractor segmentation maps. These bad-pixel masks were also used to construct a weight-map for the final stack, by summing the bad-pixel masks (in a similar manner to \citealt{vandokkum2010}).

For the UltraVISTA stacks, the ultra-deep and deep cutouts were weighted differently in the final stack as the ultra-deep stripes have an exposure time a factor $\sim10\times$ greater than the deep stripes. The images are weighted by the expected S/N gain, based on the exposure time (i.e. an image with a factor of $\sim10$ more exposure time, will result in a S/N gain of $\sim3$). The exact exposures varied between the Y, J, H and Ks bands, with the relative weights between the deep and ultra-deep also changing slightly. 

The cutouts were normalized to the sum of the flux contained in the central $1.5^{\prime\prime}\times1.5^{\prime\prime}$ (corresponding to $10\times10~\mathrm{pixel}$ for UltraVISTA images and  $25\times25~\mathrm{pixel}$ for CANDELS images). A weighted sum was performed on the masked cutouts, with the cutouts contained in the ultra-deep stripes given a heavier weight than those in the deep stripes. This summed image was divided by the weight-map to provide the final stack.  

For the UltraVISTA stacks, PSFs were generated similarly to the stacked-galaxy images. Stars within a magnitude range were chosen such that the stars had sufficiently high S/N without being saturated ($\approx16.5$ $Ks$-band magnitude). The stars were treated in the same manner as the stacks of the galaxies (i.e. normalized and averaged). To account for variations in the PSF across the mosaic, 12 different PSFs for each band were generated corresponding to 12 different regions of the mosaic ultra-deep stripes, and 9 for the deep stripes. A final PSF for the relevant band was generated from a weighted average of the 12/9 PSFs, with the weights corresponding to the number of galaxies from each field that went into the making of the stack. Thus, each stack has a uniquely generated PSF. 

For the CANDELS F160W stacks, PSFs for each of the 5 fields were taken from the 3DHST-CANDELS data release \citep{grogin2011, koekemoer2011, skelton2014}. In a similar manner to the UltraVISTA stacks, the PSF for the relevant band was generated from a weighted average of the PSF from each field, with the weights corresponding to the number of galaxies from each field that contributed to the final stack, thus each F160W stack similarly has a uniquely generated PSF. 

Fig.~\ref{fig:stacks} displays the results from the stacking analysis. Each panel contains a $24\times24^{\prime\prime}$ display of one of the UltraVISTA bands (either Y,J,H, or Ks), except the last two panels which are stacks of the CANDELS F160W data. The UltraVISTA stacks are all displayed at the same colour scale to highlight the differences in background, which increases with increasing $z$. The F160W stacks were plotted at a different colour scale for clarity, as the background is much higher.

In addition to the stacks in Fig.~\ref{fig:stacks}, 100 bootstrapped images were also generated for each stack to constrain uncertainties in the structural parameters determination (see Sec.~\ref{sec:Sersic_fitting}). Each bootstrapped image also comes with its own unique PSF that reflects the proportion of galaxies from various fields in the same manner as the original stacked images.

\subsection{Sersic Profile Fitting}
\label{sec:Sersic_fitting}


\begin{figure}
  \begin{center}
  \includegraphics[width=\linewidth]{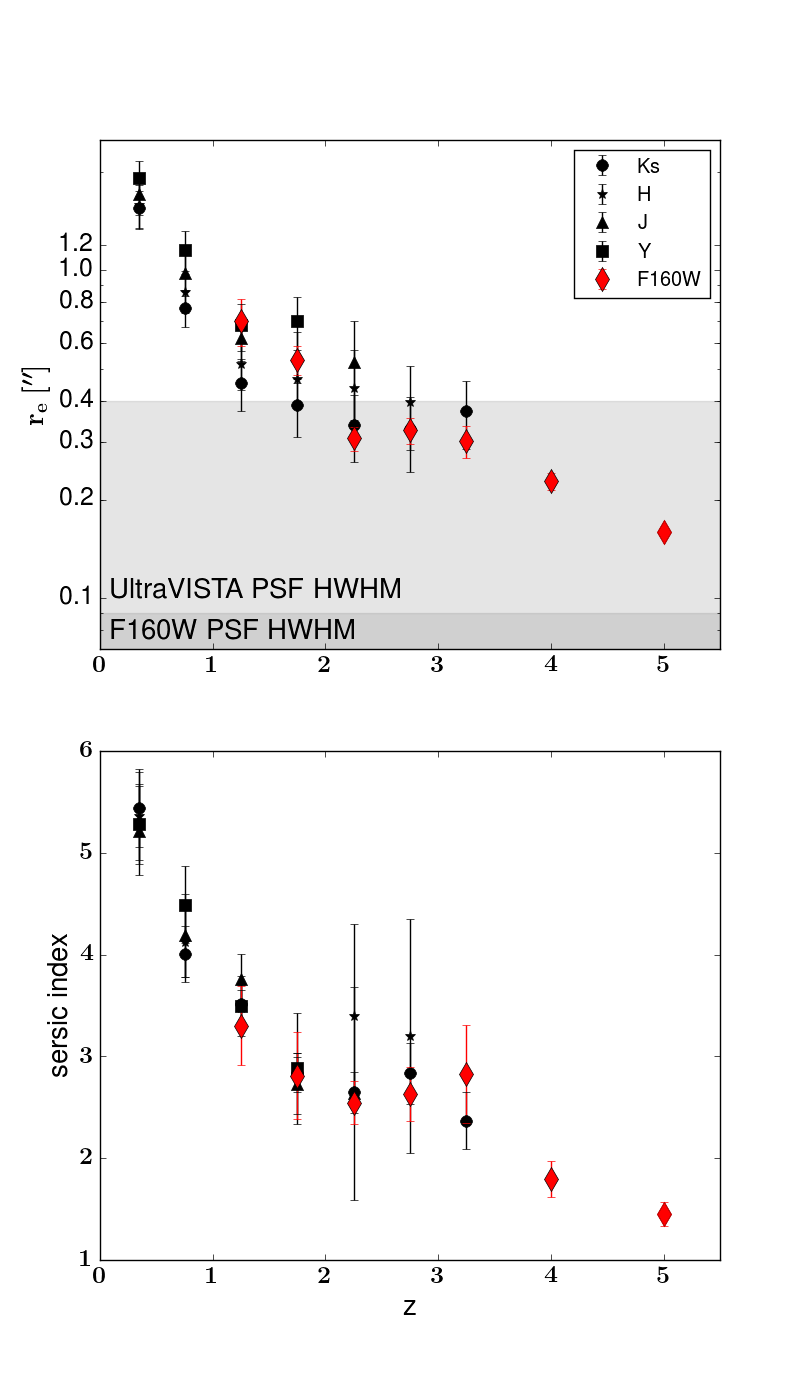}
  \caption{\textit{Top:} Best-fit effective radius in units of arc seconds as a function of $z$ for all bands. Points originating from UltraVISTA are plotted in black, with different symbols corresponding to the specific bands as indicated in the legend. HST $F160W$ are indicated by red diamonds. The seeing HWHM for both UltraVISTA and HST are displayed in light and dark grey respectively. \textit{Bottom:} Similar to the top, but with $n$ as a function of $z$. In both panels we see a progression to smaller values with $z$. In both panels, the $z=5$ point is plotted as an open symbol to remind the reader that we are mass-incomplete at that redshift.}
  \label{fig:raw_fit}
  \end{center}
\end{figure}


Sersic fitting \citep{sersic1968} of the stacked and bootstrapped images was performed using GALFIT \citep{peng2010}, with the only constraints imposed on the fits being to restrict the value of the Sersic indices to between $1<n<6$. Fig.~\ref{fig:raw_fit} shows the best fit $r_{e}$ and $n$ for each band, and each redshift bin, with the uncertainty derived from the $1\sigma$ distribution of the bootstrapped fits. The UltraVISTA derived values are in black, with each symbol corresponding to a different band. The F160W values are indicated with red diamonds, with the last symbol plotted unfilled to mark where we are incomplete. The HWHM of the UltraVISTA and CANDELS PSFs are indicated on the top panel in light-grey and dark-grey regions respectively. As seen from the top panel of Fig.~\ref{fig:raw_fit}, we resolve the stacked images to within an effective radius for UltraVISTA below $z=2$, and the $r_{e}$  is fully resolved for CANDELS in all redshift bins. Additionally, for the redshift bins for which we have stacks for CANDELS and UltraVISTA, the derived sizes and Sersic indices are roughly consistent with one another, suggesting our ground-based structural parameters are reliable. 

Absent from Fig.~\ref{fig:raw_fit} are best-fit values for $r_{e}$ and $n$ below $z=1$ for the F160W band. In these redshift bins, at the mass ranges considered, there were no galaxies present in the catalogue to contribute to a stack (see Table~\ref{tab:number}). Similarly, best-fit values for the UltraVISTA bands are not present for all redshift bins with the Y, J, H and Ks dropping out at $z=2,  2.5, 3$ and $3.5$ respectively, due to insufficient signal-to-noise in the resultant stack (and that we are incomplete in UltraVISTA at $z>3.5$).  

\subsection{Evolution in $r_{e}$}
\label{sec:re_evolution}


\begin{figure*}
  \begin{center}
  \includegraphics[width=\textwidth]{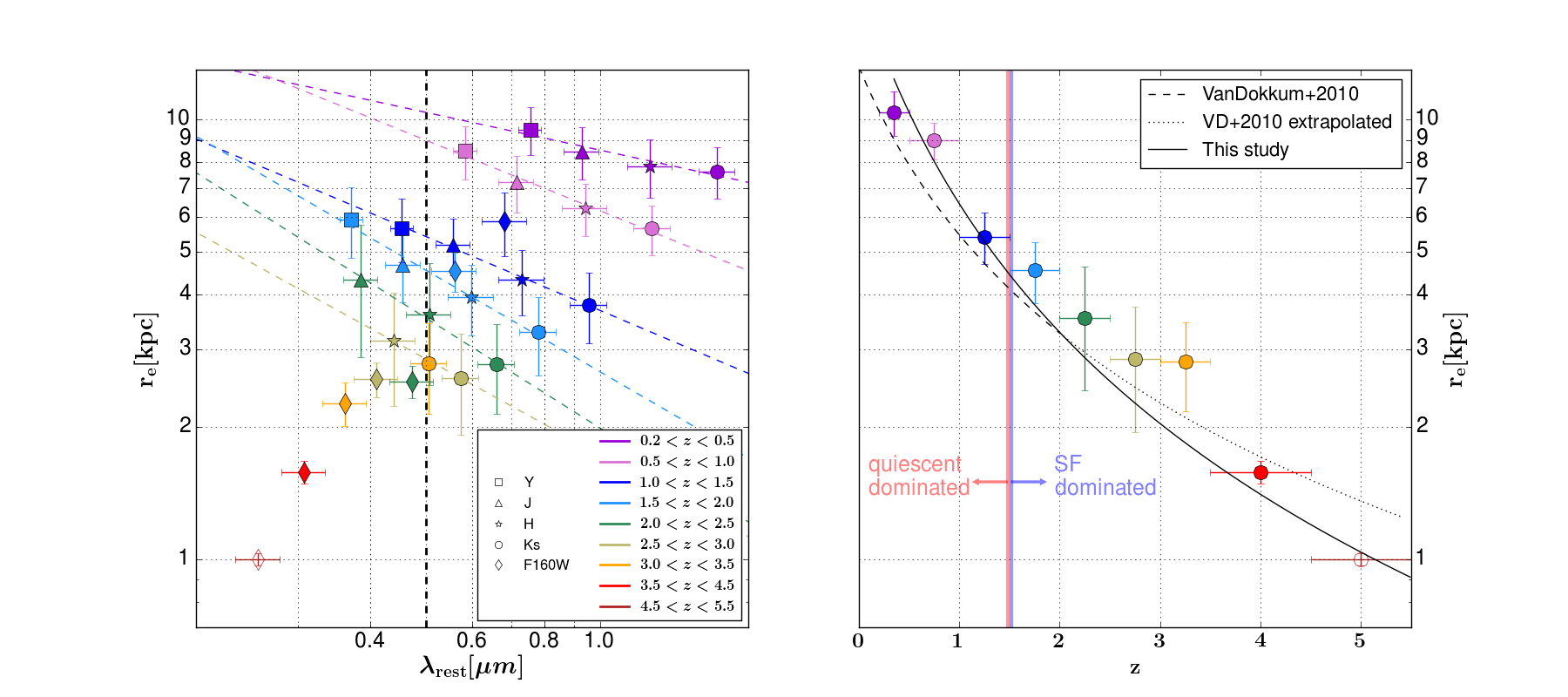}
  \caption{\textit{Left:} The effective radius plotted against the rest-frame wavelength for the stacks in all bands measured. Different shaped symbols correspond to the observed band with the same symbol convention as Fig.~\ref{fig:raw_fit}. Each color corresponds to a different redshift range, with the color convention the same as Fig.~\ref{fig:progenitors}, including plotting the $4.5<z<5.5$ symbol as open-faced to remind the reader that we are mass incomplete for that $z$-bin. Dashed coloured lines are linear best fits to the data, with the bold, black, vertical dashed line marking the rest-frame $0.5~\micron$ point, which the data at $z<3$ are extrapolated/interpolated to, so as to compare the same rest-frame sizes. $z$-ranges with only one measurement are not extrapolated for reasons discussed in Sec.~\ref{sec:re_evolution}. \textit{Right:} The size evolution of the progenitors of massive galaxies since $z\sim5$. Colored circles are the extrapolated/interpolated point at $z<3$, or the `raw' measurements at $z>3$. Over plotted are the size-$z$ relation of \citet{vandokkum2010}, as well as the size relation derived for this study.}
  \label{fig:re_rest}
  \end{center}
\end{figure*}


Due to the progression of redshift between the stacks, the $r_{e}$ and $n$ are measured at varying rest-frame wavelengths. In order to measure as closely as possible the same rest-frame wavelength, we have measured how $r_{e}$ and $n$ change with wavelength. In the left panel of Fig.~\ref{fig:re_rest} we have plotted $r_{e}$ as a function of rest-frame wavelength. Different colours correspond to different redshift bins, and different symbols demarcate the observed band (with the same symbol convention as in Fig.~\ref{fig:raw_fit}). The desired rest-frame wavelength of $0.5~\micron$ was chosen to minimize extrapolation, as well as still be red-ward of the optical-break. 

At $z<3$, we have measurements in multiple bands, and find the effective radii decrease with increasing rest-frame wavelength which is consistent with results from previous studies \citep[e.g.,][]{cassata2011, kelvin2012, vanderwel2014, lange2015}. However between $2<z<3$, the uncertainties are consistent with little to no evolution in $r_{e}$ with rest-frame wavelength. When considering the evolving properties of the progenitors with redshift, this result is also consistent with the literature. \citet{vanderwel2014} who measured the sizes of galaxies from CANDELS  at $0<z<3$, found the size-gradient with rest-frame wavelength was steepest for galaxies at high-mass, and low-redshift, and flatter for low-mass galaxies. As the progenitors decrease in mass with redshift, we expect a flattening of this gradient. The difference in size-gradients is also seen in local populations. \citet{kelvin2012} found size-gradients to be flatter for late-type galaxies in the GAMA survey. Because we only have measurements in one band for $z>3.5$, and we are dominated by late-type galaxies at high redshift, we have not extrapolated $r_{e}$ between $3.5<z<5.5$ and assume the measurement is representative of the $r_{e}$ at $0.5~\micron$. This is assuming that the size-gradient will be flat for low-mass, late-type galaxies at high redshift.

In the right panel of Fig.~\ref{fig:re_rest}, we have plotted $r_{e}$ at $0.5~\micron$ as a function of redshift. It is clear from both panels of Fig.~\ref{fig:re_rest} that the $r_{e}$ decreases out to $z=5$ which is consistent with previous results using a diverse set of methods to select progenitors \citep[e.g.,][]{vandokkum2010, williams2010, damjanov2011, mosleh2011, oser2012, barro2013, patel2013a, straatman2015, ownsworth2016}. In spite of the different choice of progenitor selection, below $z=2$ our measurements fall broadly on the same relation found by \citet{vandokkum2010} (our values are systematically larger, but this is likely a reflection of our slightly higher mass selection). This result is not surprising, and the consistency is reflected in the right panel of Fig.~\ref{fig:progenitors}, where at $z<2$, the mass of the progenitors chosen using a fixed vs. evolving cumulative number density are within the uncertainties in both the mass function and the semi-analytic models. Although we measure a slightly steeper relation than \citet{vandokkum2010}, it is surprising how well the relation is extrapolated at $z>2$ given that we are selecting galaxies which are distinct in mass from the fixed-cumulative number density selection. 

\begin{figure}
  \begin{center}
  \includegraphics[width=\linewidth]{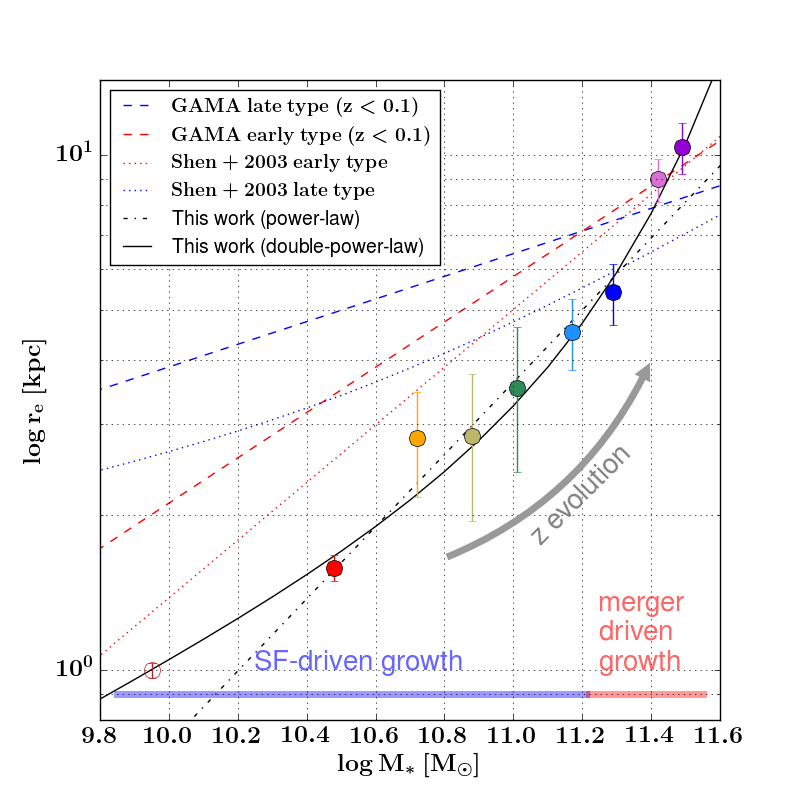}
  \caption{The implied mass-size evolution of the progenitors of massive galaxies. Circles are measurements from the stacks from the present study, with each point representing a different redshift. The $r_{e}$ plotted above are the same values taken from the right-panel of Fig.~\ref{fig:re_rest}. Symbol color plotting convention is the same as Fig.~\ref{fig:progenitors}, with the lowest-$z$ points corresponding to the most massive galaxies, and monotonically decreasing to the highest-$z$. We have plotted the highest-$z$ point as an open face symbol to remind the reader we are mass incomplete for that $z$-bin. Plotted above are the $g$-band local mass-size relations for late (dashed blue line) and early type (dashed red line) galaxies from the $GAMA$ survey \citep{lange2015}, as well as the mass-size relations from \citet{shen2003}. For the lowest 2 $z$ bins (i.e. $z<1$), our galaxies fall precisely on the local mass-size relation for early type galaxies, but are systematically below the relations at higher $z$. Also plotted above are the best fit single and double power-law relations to our data.}
  \label{fig:mass_size}
  \end{center}
\end{figure}

In Fig.~\ref{fig:mass_size} we investigate the evolution of the mass-size plane. We have taken the values of $r_{e}$ from the right panel of Fig.~\ref{fig:re_rest}, and plotted them against their respective progenitor masses, with the highest mass associated with the lowest redshift bin. For comparison purposes, we have over-plotted the mass-size relations from \citet{shen2003} and \citet{lange2015} for both early and late type galaxies. For \citet{lange2015}, who investigate the mass-size relations as a function of rest-frame wavelength, we use their $g$-band relations which corresponds most closely to a rest-frame wavelength of $0.5~\mathrm{\micron}$. The measured $r_{e}$ from our stacking analysis fall on the SDSS and GAMA mass-size relations for early-type galaxies at $z<0.1$. However for all other redshift bins, our galaxies fall below the local-mass size relation, consistent with \citet{vandokkum2010}. Also plotted in Fig.~\ref{fig:mass_size} are simple single (dash-dot line) and double (solid line) power-law fits of the form 

\begin{equation}
r_{e} = aM_{*}^{b}
\label{eq:single_pl}
\end{equation}

\begin{equation}
r_{e} = \gamma M_{*}^{\alpha}\left(1+ \frac{M_{*}}{M_{0}}\right)^{\alpha-\beta}
\label{eq:double_pl}
\end{equation}

From Fig.~\ref{fig:mass_size}, we see that the double power-law is a more appropriate fit for our data, with the parameters $\gamma=2.9\times10^{-4}$, $\alpha=0.35$, $\beta=2.1\times10^{3}$ and $\log(M_0/M_{\odot})=14.77$. Continuing with the plotting convention in previous figures, our $4.5<z<5.5$ point has been plotted as an open-face symbol to highlight incompleteness issues within that bin. It is interesting to note that this point has not been included in any of the power-law fits, and the fact that it falls on on the extrapolation of the double power-law is not designed. 

The evolution in the mass-size relation in Fig.~\ref{fig:mass_size} can be broadly separated into two phases. At $z>1.5$, the mass-size evolution is relatively linear (in log-log space), with most points falling along a single power-law. At $z<1.5$, the size growth becomes more efficient, and no longer follows the same single power law as before. This is broadly consistent with patterns we have seen in Fig.~\ref{fig:uvj_average} and Fig.~\ref{fig:re_rest} i.e., that star-formation and mergers are dominating mass and size growth at different epochs, with this changeover occurring at $z\sim1-2$. Before this time, mass was primarily added via star formation, which has been shown to be ineffective at altering the structure of massive galaxies \citep{ownsworth2012}. At these redshifts, we see a marked increase in the quiescent fraction of the progenitors. As star formation is no longer an available pathway to mass growth, the growth is dominated by minor mergers, which efficiently increases the $r_{e}$ (see Sec.~\ref{sec:intro} and references therein). 

\subsection{Evolution in $n$}
\label{sec:n_evolution}

\begin{figure*}
  \begin{center}
  \includegraphics[width=\textwidth]{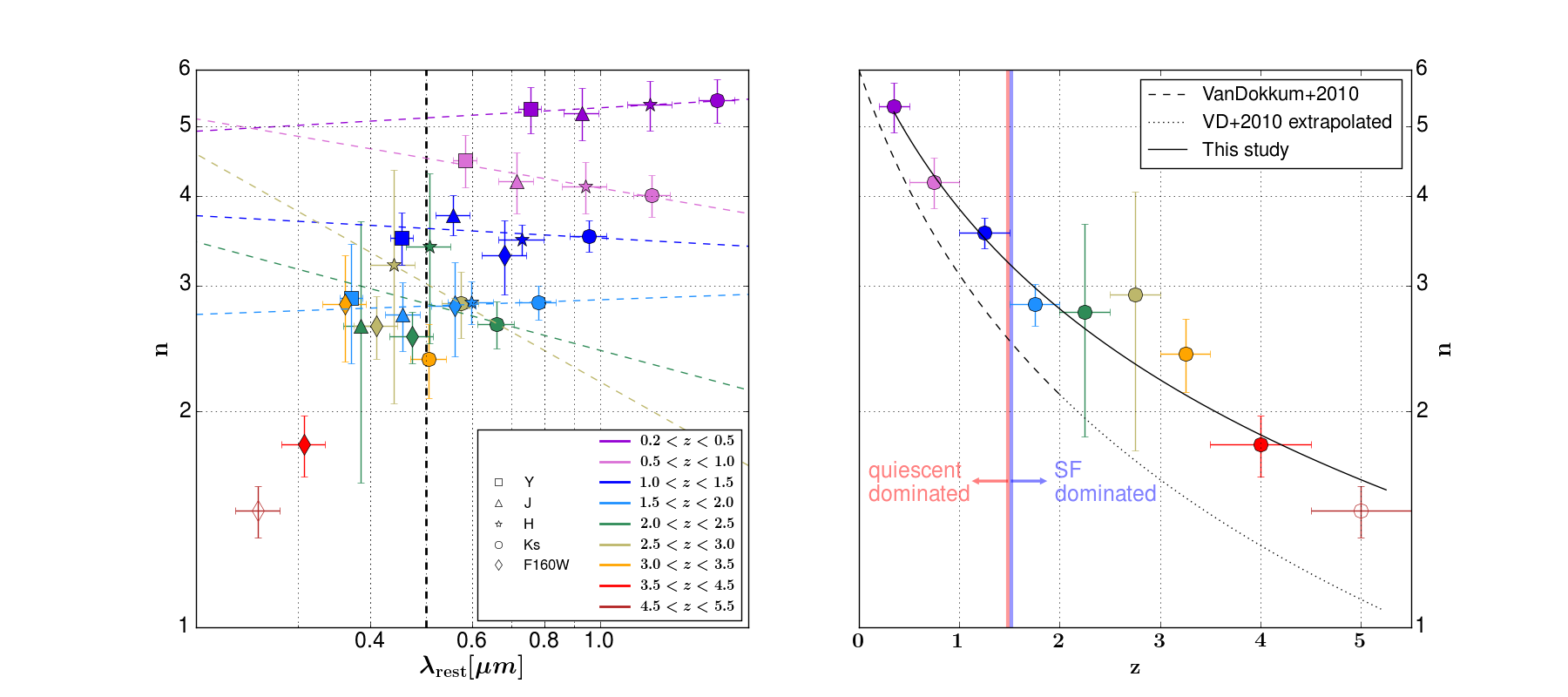}
  \caption{The same figure as Fig.~\ref{fig:re_rest}, but with Sersic index instead of the effective radius \textit{Left:} Sersic index plotted against the rest-frame wavelength with the same plotting convention as Fig.~\ref{fig:re_rest}. As the relation between $n$ and $\lambda_{rest}$ is consistent with flat, there is no extrapolation to the $0.5~\micron$ point. Instead, the values are averaged to produce a representative $n$ for each $z$-bin. \textit{Right:} The evolution of $n$ with $z$. Over-plotted are the best fit relation for this study, as well as the relation from \citet{vandokkum2010}.}
  \label{fig:n_rest}
  \end{center}
\end{figure*}

Fig.~\ref{fig:n_rest} is analogous to Fig.~\ref{fig:re_rest}, except we investigate how the Sersic index $n$ changes with rest-frame wavelength in place of $r_{e}$. From the left panel of Fig.~\ref{fig:n_rest} we see little to no evolution in $n$ with wavelength at any redshift. We have therefore taken an average $n$ weighted by the bootstrapped uncertainty in each band to measure a representative $n$ for each redshift bin. At $z>3$ where we only have one measurement for each stack, the measurement was considered representative. The resulting values are plotted in the right panel of Fig.~\ref{fig:n_rest}. 

Fig.~\ref{fig:n_rest} shows a clear downward trend of $n$ with redshift, consistent with previous findings out $z=2$ \citep[e.g.,][]{vandokkum2010}. This trend is also expected given the evolution in the quiescent fraction. Actively star forming galaxies tend to have lower $n$/are more centrally concentrated than their quiescent counterparts \citep[e.g.,][]{lee2013, freeman1970,lange2015,mortlock2015}, thus at $z>1$ the decrease in $n$ is likely driven by morphological changes between each redshift bin which we also see reflected in the evolution of the mass-size relations (Fig.~\ref{fig:mass_size}). \citet{vandokkum2010} also found $n$ to decrease with redshift, although their relation is steeper than the one measured in the current study. However the $n-z$ relation from \citet{vandokkum2010} was derived from galaxies at $z<2$, where the slopes are comparable, but where we measure systematically higher $n$. 

\subsection{Mass Assembly}
\label{sec:assembly}

\begin{figure*}
  \begin{center}
  \includegraphics[width=\textwidth]{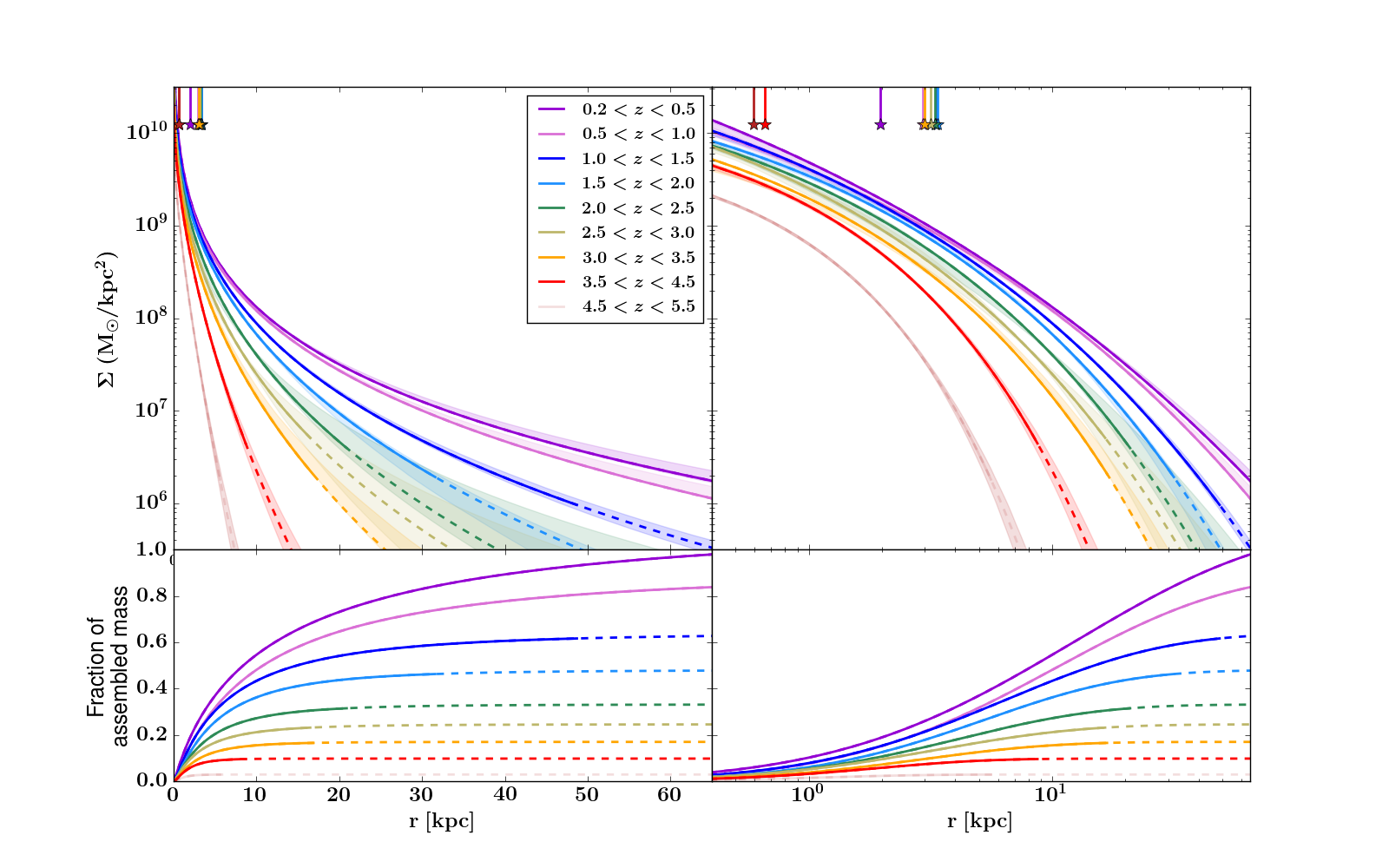}
  \caption{\textit{Top:} The projected surface mass-density profiles for our stacks (presented in both log-linear and log-log scales). Each profile is a Sersic, with the $r_{e}$ and $n$ taken from the right panels of Fig.~\ref{fig:re_rest} and Fig.~\ref{fig:n_rest}, with the constraint on normalization that the integrated mass within $75~\mathrm{kpc}$ be equal to the implied progenitor mass from Fig.~\ref{fig:progenitors}. The faded filled region corresponds to profiles within the $16^{th}$ and $84^{th}$ percentile from the bootstrapped images. The transition from solid to dashed line in the profile marks the point where the error in the profile is at the level of the background. The PSF HWHM for each redshift is also marked with a vertical line ending in a star at the top of each plot. \textit{Bottom:} The fraction of assembled mass with radius for each profile (presented in both log-linear and log-log scales). The curves are all normalized to the total mass at $0.2<z<0.5$. The curve for $4.5<z<5.5$ is faded to remind the reader that we are mass incomplete in that $z$-bin.}
  \label{fig:sb}
  \end{center}
\end{figure*}

Equipped with measurements of $r_{e}$ and $n$, we can investigate surface-density profiles, and mass assembly as a function of radius. To generate these profiles, we have assumed that the mass-to-light ratio is constant across the profile, and that all the mass can be found within a radius of $75~\mathrm{kpc}$. Given these assumptions, and that the integrated mass within $75~\mathrm{kpc}$ must equal the total mass found in the right panel of Fig~\ref{fig:progenitors} (i.e., the same constraints used in \citealt{vandokkum2010}), we have generated stellar-mass density profiles which can be found in Fig.~\ref{fig:sb}. 

Fig.~\ref{fig:sb} shows the Sersic fits using the values of $r_{e}$ and $n$ for each redshift bin in the right panels of Fig.~\ref{fig:re_rest} and Fig.~\ref{fig:n_rest} respectively. The transition between the solid and dashed lines for each profile marks the point when the error in the background becomes significant. Since many of the values of $r_{e}$ and $n$ are either interpolated, or averaged (see Sec.~\ref{sec:re_evolution} and Sec.~\ref{sec:n_evolution}), the profiles from which this transition point was determined were the closest to the rest-frame wavelength of $0.5~\micron$ (these are the same bands which are displayed in Fig.~\ref{fig:stacks}). 

Fig.~\ref{fig:sb} illustrates that the majority of mass build-up in galaxies since $z=4$ occurs in the outskirts, consistent with previous findings and the inside-out growth paradigm for massive galaxies \citep{vandokkum2010, toft2012, bezanson2013b, vandesande2013, belli2014a, margalef2016}. It is only at $z=5$ that we begin to see significant growth in the inner regions. Important to note is that as we are incomplete in that redshift bin, we will be biased towards blue, and possibly diskier galaxies which would likely have lower values of $n$.  However given the trend of Sersic index with redshift found in Fig.~\ref{fig:n_rest}, this does not seem to be an unreasonable depiction of the progenitors.

\begin{figure}
  \begin{center}
  \includegraphics[width=\linewidth]{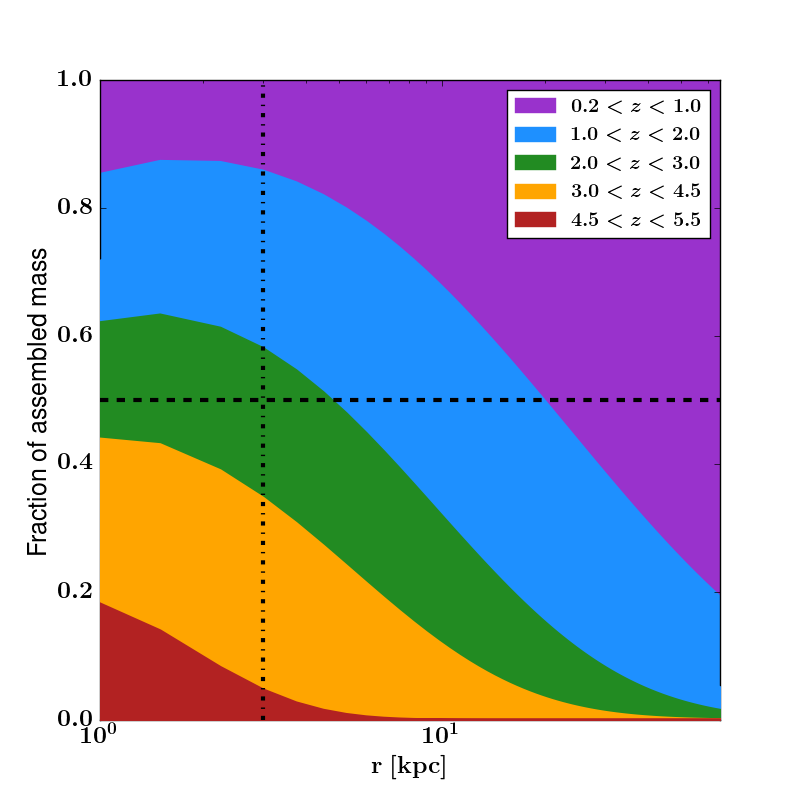}
  \caption{The fractional build-up of stellar mass, as a function of radius, assembled at various $z$ intervals (i.e. each mass profile in Fig.~\ref{fig:sb} divided by the mass profile for $0.2<z<0.5$). The horizontal dashed lines markes the 50\% assembly point, and the vertical dot-dashed line is drawn at the $3~\mathrm{kpc}$ point for clarity. From this plot, the formation redshift for the interior vs. exterior regions can be seen, with the inner regions containing 50\% of their final stellar mass between $2.0<z<3.0$, with the outer region $z$ of formation lagging behind.}
  \label{fig:mass_assembly_r}
  \end{center}
\end{figure}

In Fig.~\ref{fig:mass_assembly_r}, we have divided the surface mass density profile for each redshift bin from Fig.~\ref{fig:sb} by the surface mass density profile at $0.2<z<0.5$. In this way, we are able to trace the fractional mass assembly as a function of radius. At the highest redshift bins, we see the central regions are the first to form, with very little of the stellar mass beyond $3~\mathrm{kpc}$ in place at $z\sim5$. Between $3.0<z<4.5$, we see rapid growth, with the fraction of mass assembled in the inner regions more than doubling. It is also in this redshift interval that a not insignificant fraction of stellar mass is assembled between $3$ and $10~\mathrm{kpc}$. We can trace the redshift of formation as a function of radius by tracing the horizontal dashed-line in Fig.~\ref{fig:mass_assembly_r}, which marks the point at which half of the stellar mass was assembled. As you trace from small to large radii, the dashed line crosses different coloured regions, indicating that the interior regions were the first to assemble, with the outer regions assembling at later and later times, indicative of `inside-out' growth.


We can trace this growth quantitatively by considering the total mass in and outside the $3~\mathrm{kpc}$ boundary. We have de-projected the surface density profiles of Fig.~\ref{fig:sb}, and separated the mass growth into stellar mass assembly that is within $r<3~\mathrm{kpc}$, and exterior to $r>3~\mathrm{kpc}$. The total mass assembly is indicated in black, and is the same mass assembly seen in the right panel of Fig.~\ref{fig:progenitors}. From the red line, we see continuous, albeit decelerating, mass assembly from $z=5$ to $z=0$. This is inconsistent with previous works such as \citet{vandokkum2010} and \citet{patel2013a} who found the interior regions are consistent with no assembly since $z=2$, oft cited to be evidence of `inside-out' growth, although it depends on precisely what is meant by this term.


\begin{figure}
  \begin{center}
  \includegraphics[width=\linewidth]{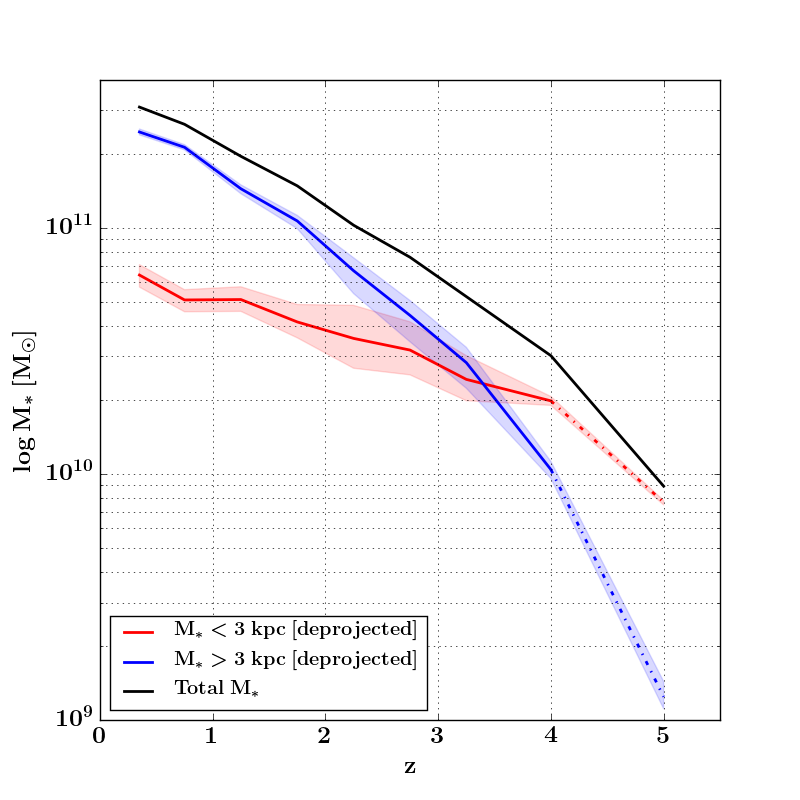}
  \caption{The total projected mass within $3~\mathrm{kpc}$ (red-line) and outside $3~\mathrm{kpc}$ as implied by integrating the profiles from Fig.~\ref{fig:sb}. The last symbol is plotted as open faced to remind the reader that we are mass incomplete in that $z$-bin. There is growth in both radial regions, however the growth is not self-similar with the growth outside $r=3~\mathrm{kpc}$ proceeding at a faster pace than the inner regions.}
  \label{fig:mass_assembly}
  \end{center}
\end{figure}


It is important to note from Fig.~\ref{fig:mass_assembly_r} and Fig.~\ref{fig:mass_assembly} that even though the regions outside $3~\mathrm{kpc}$ experience a greater growth rate than the inner regions, there is still significant mass build-up from $z=5$ to $z=0$ in the interior. Although the growth between the inner and outer regions is not self-similar, the growth is not necessarily `inside-out' as described in previous works \citep[e.g.,][]{vandokkum2010, vandesande2013}, especially when considering the mass assembly at $z>3$. At these redshifts, significant stellar mass is assembled at all radii (although mass accretion is concentrated in the central regions). 

\subsection{Comparisons with simulations}

There have been many comparisons between the mass growth of galaxies in extra-galactic surveys (i.e. mass functions) to hydrodynamical galaxy simulations \citep[e.g.,][]{vogelsberger2014, schaye2015}. In fact, the EAGLE simulation has been calibrated to reproduce the galaxy stellar mass function at $z=0$. However, there remain few examples \citep[e.g,][]{snyder2015,wellons2015} in the literature which explicitly compare the evolution of structure in simulations to observations . In this section, we endeavour to make such a comparison.


\begin{figure}
  \begin{center}
  \includegraphics[width=\linewidth]{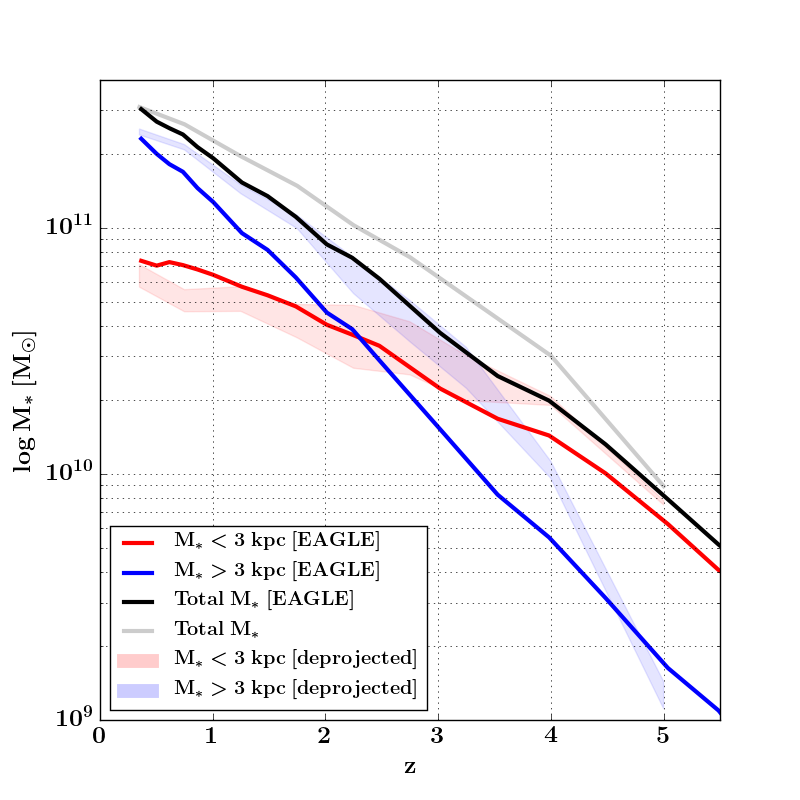}
  \caption{Above show the the build-up of stellar mass inside (red line) and outside (blue line) a $3~\mathrm{kpc}$ aperture as predicted by the EAGLE simulation, as well as the total stellar mass evolution (black line). The faded colours is the mass evolution from this study, with the colours corresponding to the same regions as the simulations. The simulations show rapid build up of the outer regions, which is qualitatively similar to the data. The main difference between the observations and the simulations is the total mass evolution proceeds more rapidly in the simulations, with most of the effects see in the build up of the outer regions.}
  \label{fig:mass_assembly_eagle}
  \end{center}
\end{figure}


In Fig~\ref{fig:mass_assembly_eagle}, we see how the mass assembly as implied by our observations compares to the EAGLE simulation \citep{schaye2015}.  In Fig.~\ref{fig:mass_assembly_eagle}, we see the total mass assembly (black line), the mass assembly within a $3~\mathrm{kpc}$ aperture and the mass assembly outside a $3~\mathrm{kpc}$ aperture. Also plotted in Fig.~\ref{fig:mass_assembly_eagle} are the de-projected aperture masses from this study for comparison. The progenitors in EAGLE are defined as the `true' progenitors, and are selected in a similar method to the dark-matter halo mergers trees from \citet{behroozi2013} which inform the abundance matching technique, i.e. only the most massive progenitor from the precursors of a merger is considered. The progenitors were traced from all galaxies within the EAGLE simulation that have a stellar mass within $0.1~\mathrm{dex}$ of $\log(M_{*}/M_{\odot})\sim11.5$ (i.e. chosen to match the starting point of this study), which amounted to 24 galaxies. The aperture masses from EAGLE quoted above are averages from the progenitors of these 24 galaxies. 

A qualitative comparison between the simulations and the observations show remarkable agreement. For the mass within $3~\mathrm{kpc}$, the agreement is always within a factor of 2, which is within the uncertainty associated with the assumptions made when determining stellar masses from photometry \citep{conroy2009}. Both methods predict the same overall trend i.e. that there is a steady build up of stellar mass within $3~\mathrm{kpc}$, and rapid assembly at later times at radii larger than $3~\mathrm{kpc}$. The main difference between the simulations and observations is that EAGLE predicts a more rapid assembly of the progenitors. The progenitors in EAGLE must assemble more mass in the same period of time in order to result in the final stellar mass of $\mathrm{\log{M_{*}/M_{\odot}}}=11.5$ at $z\sim0.3$. This offset is not entirely unexpected, given differences between the evolution of the observed and simulated galaxy stellar mass functions at high-$z$ in the mass ranges considered for this study ($\sim10^{10}-10^{11}~\mathrm{M_{\odot}}$, \citealt{furlong2015}). 

The progenitors in EAGLE must assemble more mass in the same period of time in order to come to the same descendant mass by $z\sim0.3$; and given the agreement with observations at $r<3~\mathrm{kpc}$, nearly all of this mass growth must occur in the outer regions. This suggests the progenitors in EAGLE are more centrally concentrated than observed, except at $z>4$. Between $4<z<5$, the fraction of stellar mass outside a $3~\mathrm{kpc}$ aperture is broad agreement with the observations, which does not follow the trend at $z<4$. One possible reason for this is the effective radius at these redshifts is close to $1~\mathrm{kpc}$, which suggests nearly all of the total bound mass in the galaxy would be within $3~\mathrm{kpc}$, which is not true at lower redshifts. 

Some caveats that could affect the above comparison are some assumptions that were made in the observations, in particular the assumption of a constant mass-to-light ratio for our surface mass density profiles. If there is a strong gradient of stellar age with radius in the progenitors, and the interiors are older (which would be consistent with what we see in Fig.~\ref{fig:mass_assembly_r}), then we would over-predict the fraction of the total stellar mass which is located at large radii, bringing us closer to agreement with the simulations. A similar effect would be expected if there are also strong gradients in dust. An analysis of forthcoming virtual observations from EAGLE with the effects of dust and inter-cluster light taken into account would be a better comparison, the investigation of which is beyond the scope of this paper.

\section{Discussions and Conclusions}

\subsection{Mass and size growth at $z<2$}

In this paper, we have selected the progenitors of today's massive galaxies through an evolving cumulative number density technique, and have made image stacks to infer their evolution with redshift. Based on rest-frame $U-V$ and $V-J$ colours, we find the progenitors of massive galaxies become increasingly star forming out to higher redshift, and by assuming Sersic profiles for the mass distribution, we find the progenitors decrease in both $r_{e}$ and $n$. These trends are qualitatively similar to previous studies which select based on fixed  \citep[e.g.,][]{vandokkum2010, patel2013a, ownsworth2014}, and evolving \citep{marchesini2014} cumulative number densities at $z\lesssim2$

Although the qualitative trends are consistent with the literature, there are quantitative differences, especially in regards to the evolution of the central mass densities with redshift. Previous works \citep[e.g,][]{vandokkum2010, patel2013a, vandokkum2014} have found little to no mass assembly in the inner regions  ($r<\mathrm{2~kpc}$), and find that mass assembly occurs in an inside-out fashion with the majority of mass growth since $z\sim2$ occurring at $r>\mathrm{2~kpc}$ (although in \citealt{vandokkum2010} at $\sim1~\mathrm{kpc}$, there is a spread in mass-density of at least $0.1~\mathrm{dex}$ since $z=2$ suggesting modest mass growth). In this study, we find the central regions have accumulated  $\approx50\%$ of their mass between $2.0<z<2.5$, but continue to experience mass growth out to $z=0.2$, albeit at a lower rate (i.e. we find $\sim90\%$ of the mass within $2~\mathrm{kpc}$ was in place by $z\sim1$).

The suspected cause of this discrepancy is the differences which arise between a fixed vs. evolving cumulative number density selection. By using a fixed cumulative number density selection, one is biased towards the most massive progenitors \citep[e.g.,][]{clauwens2016, wellons2016b}. This is a result of the fact that an abundance matching technique (i.e \citealt{behroozi2013}) predicts higher number densities with increasing redshift, whereas a fixed cumulative number density will select galaxies at a steeper point in the mass function which is inhabited by higher mass galaxies. We have tested this hypothesis by re-measuring the surface mass density profiles for a fixed cumulative number density selection (see Fig.~\ref{fig:progenitors} for the mass assembly history), and do find the redshift evolution in central regions of the stellar surface mass density profiles is considerably weaker than for an evolving number density selection (Fig.~\ref{fig:sb_fixed}). Details of this analysis can be found in an attached appendix.

In contrast to a fixed cumulative number density selection, \citet{vandokkum2014} selected galaxies based on their stellar surface mass density within $1~\mathrm{kpc}$ (i.e. 'dense cores'), and found evidence that the interiors are formed first, with the outer radii forming around them. This inconsistency can also be attributed to selection, and the progenitors \citet{vandokkum2014} select are likely a subpopulation of the progenitors of massive galaxies. Since they are selected on central stellar density, and central stellar density is correlated with quiescence, they will not select star forming progenitors. This is evidenced by the differences in quiescent fraction at $2.0<z<2.5$; \citet{vandokkum2014} find a quiescent fraction of $57\%$, whereas the selection of the current study has a quiescent fraction of $23\%$ in the same redshift range. 

The most massive progenitors are likely to host older stellar populations, have less star formation, and more compact configurations due to rapid early assembly . As these progenitors would have assembled first, they experience more passive evolution in their central regions between $z=2$ and today \citep[e.g.,][]{vandesande2013}. The star forming progenitors however, still must quench, and might involve more violent events, such as disk instabilities which result in compaction, i.e. the driving of mass towards smaller radii \citep{dekel2014, barro2014}. By averaging these populations, one would expect modest gains in stellar mass density in the central regions, which is what is seen in our analysis. 

An important caveat to consider when selecting progenitors at systematically higher number densities is the effect of a lower normalization to the mass profiles. Our profiles are designed such that $100\%$ of the stellar mass, as determined from the mass functions as outlined in Fig.~\ref{fig:progenitors}, is contained within $75~\mathrm{kpc}$. If at each $z$ step we have a slightly lower mass selection than studies based on a fixed cumulative number density selection, the normalization of the profile will trend to lower values which imposes sustained mass growth in the central regions (see Fig.~\ref{fig:sb_fixed}, and discussion in the appendix).

Although we find the progenitors continue to assemble mass at all radii, the growth rate at small and large radii is not self-similar. The fractional growth rate is higher at larger radii, consistent with the idea that minor mergers play a dominant role in the mass assembly at $z<1.5$, and especially at $z<1$ as found by \citet{newman2012, whitaker2012, belli2014b, belli2015, vulcani2016}.

This is also in agreement with our quiescent fractions, which are $>90\%$ at $z<1$, suggesting that the majority of the mass growth cannot be from star formation. However between $1<z<2$ our star-forming fraction exceeds $50\%$, suggesting the increasing importance of star formation in mass assembly, which is in broad agreement with \citet{vulcani2016} who find star-formation and minor mergers play equal roles in mass growth during this epoch. Additionally, $H{\alpha}$ maps of massive star-forming galaxies between $0.7<z<1.5$ reveal that the disk scale lengths are larger in $H{\alpha}$ than in the stellar continuum, suggesting that star formation also contributes to the mass build-up at large radii \citep{nelson2016}, and not just in the inner regions.

\subsection{Mass and size growth at $z>2$}

In addition to comparisons with other works, which are largely limited to $z<2$, we have selected progenitors, and generated stacks for galaxies out to $z=5.5$. In this regime we see a continuation of the trends at $z<2$, i.e. progenitors are smaller, and have Sersic indices which imply more disk-like configurations than spheroidal. This is consistent with the evolution of our quiescent fraction which continues to decrease with increasing $z$, suggesting the progenitors are dominated by star forming galaxies which also tend to have disk-like morphology, which is observed in massive galaxies at high redshift \citep[e.g.,][]{vanderwel2011, wuyts2011, bruce2012, newman2015}. This is in agreement with the prediction of \citet{patel2013a} who posited that the progenitors of massive galaxies at $z>3$ will continue the trend towards smaller sizes. 

The trends in the evolution of the mass-size relation, the $r_{e}$, the Sersic index, the $UVJ$ colour evolution, and the FIR derived SFRs all corroborate the idea that $z\sim1.5$ represents a transitional period in how the progenitors of massive galaxies assemble their mass. At $z>1.5$ the $UVJ$ colours and the FIR SFRs suggests the progenitors are actively forming stars, and the Sersic index suggests those stars are consistent with being distributed in an exponential disk. The change in power-law slope at $z\sim1.5$ in the evolution of the mass-size plane suggests a change in assembly method; one in which the size evolves more efficiently with mass than at higher redshift, consistent with the minor merger scenario (see Sec.~\ref{sec:intro} and references therein). This is further corroborated by the fact that the FIR SFR is insufficient to account for the rate of stellar mass assembly at $z<1.5$ (Fig.~\ref{fig:sfr}).

This study supports the scenario that the progenitors of massive galaxies begin with a disk-like morphology with the disk forming concurrently with the central regions (i.e. the `bulge'). At some point, the disk morphology is destroyed, either by major mergers, or disk instabilities which may also be responsible for the increase in quiescent fraction. Evidence of disks \citep[e.g.,][]{vandokkum2008, vanderwel2011, wuyts2011, bruce2012, bell2012} and rotation \citep{newman2015} in massive compact quenched galaxies are seen at intermediate ($1.5<z<3$) redshifts  which confirms that at least \textit{some} of the massive progenitors host/hosted disk-like morphology. By $z=1.5$, assembly is less violent, with mass growth dominated by minor mergers, and more passive quenching (i.e. gas exhaustion) until $z=0$. 

The scenario that the progenitors of massive galaxies begin as disks have support in cosmological simulations. \citet{fiacconi2016b} simulated the assembly of the main progenitor of a $z=0$ ultra-massive elliptical, and found the progenitor to be disk dominated, with an exponential brightness profile at $z>6$ which had experienced several major mergers at $z>9$. The `survival', or more accurately, the reassembly of the disk after a major merger is feasible, provided the major mergers are sufficiently gas rich \citep[e.g.,][]{hopkins2009a}. \citet{fiacconi2016b} also calculated the Toomre parameter for their simulated disk and found it to be stable against fragmentation for all resolved spatial scales, with the disk supported by a turbulent inter-stellar medium thought to be due to feedback from star-formation. They also predict, that gas-rich star forming disks at $z>5$ should not host a significant bulge, but is rather built up by mergers occurring at $2<z<4$ \citep{fiacconi2016a}. This is consistent with our analysis which show the majority of the stellar mass in the central regions (i.e. $r < 1~\mathrm{kpc}$, which we take as a proxy for the bulge) is assembled between $2.0<z<5.5$. 

Stacking analysis is a useful tool to probe the average properties of low-surface brightness features of a population of galaxies. However, specific aspects of the morphology are lost in a stack. To verify our hypothesis about the nature of the progenitors of today's massive galaxies will require resolution and sensitivity of spaced based observatories such as HST. At high-$z$, the rest-frame optical emission is shifted further into the infrared, of which future space observatories such as JWST which will observe at wavelengths beyond the $K$-band, will prove to be invaluable in determining the nature of 'regular' galaxies at $z>2$.

\section{Summary}

To briefly summarize the paper, we have traced the stellar mass evolution of the median progenitors of $\log{M_{*}/M_{\odot}}=11.5$ galaxies at $z=0.35$ using abundance matching techniques. Using photometric data from the UltraVISTA and 3DHST surveys and their associated catalogs, we have used stacking analysis to trace the mass assembly of the progenitors out to $z=5.5$. By fitting the images stacks with 2D convolved Sersic profiles, we have found the following.

\begin{enumerate}
	\item Selecting progenitors based on an evolving cumulative number density selection results in progenitors that are less massive than if selected based on a fixed cumulative number density selection. This discrepancy becomes significant at $z>2$.  
	\item The progenitors of massive galaxies become progressively more star forming, with star forming fractions exceeding 50\% at $z>1.5$ as determined by their rest-frame $U-V$ and $V-J$ colours. 
	\item The progenitors decrease in both effective radius and Sersic index with increasing redshift, which is consistent with the picture that the progenitors of today's massive galaxies began with disk-like morphology.
	\item The progenitors continue to assemble mass at all radii until $z=0.35$, which suggests a more complex mass assembly then `inside-out' growth. 
	\item Even though galaxies continue to assemble mass in their interiors to low redshift, the redshift at which half of the resultant stellar mass is assembled is higher for the interiors than the exterior regions, with $z_{f,r=3~kpc}\sim2-3$, and $z_{f,r=10~kpc}\sim1-2$.
	\item A brief comparison between the implied mass assembly of this study to results from the EAGLE simulation show a very similar qualitative trend. However the results from simulations imply a more rapid assembly of the outer regions. 
\end{enumerate}

\section{Acknowledgements} 

The authors would like to thank the anonymous referee, as well as Matthew Ashby, and Joop Schaye for the helpful comments and discussion which greatly improved this paper. The research leading to these results has received funding from the European Research Council under the European Union's Seventh Framework Program (FP7/2007-2013)/ERC Grant agreement no. EGGS-278202.  This work is based on data products from observations made with ESO Telescopes at the La Silla Paranal Observatory under ESO programme ID 179.A-2005 and on data products produced by TERAPIX and the Cambridge Astronomy Survey Unit on behalf of the UltraVISTA consortium. This work is also made possible by the observations taken by the 3D-HST Treasury Program (GO 12177 and 12328) with the NASA/ESA HST, which is operated by the Association of University for Research in Astronomy, Inc., under NASA contract NAS5-26555. DM acknowledges the support of the Research Corporation for Science Advancement's Cottrell Scholarship, and the National Science Foundation under Grant No. 1513473. KIC acknowledges funding from the European Research Council through the award of the Consolidator Grant ID 681627-BUILDUP. This research has made use of NASA's Astrophysics Data System. 

\bibliographystyle{apj}
\bibliography{lit}

\appendix
\section{ The effects of a fixed cumulative number density selection on the stellar surface mass density profiles}

In this appendix, we briefly explore the effects of mass selection on the stellar surface mass density profiles. A key finding of this study is that the central ($r<1-2~\mathrm{kpc}$) stellar surface mass densities evolve more strongly than observed in earlier works \citep[e.g.,][]{vandokkum2010, patel2013a, vandokkum2014}. It was suspected that this discrepancy was a result of the different number density selections (i.e. a fixed vs. evolving cumulative number density selection as discussed in Sec.~\ref{sec:selection}), with a fixed number density selection yielding more massive progenitors (Fig.~\ref{fig:progenitors}). 

\begin{figure*}
  \begin{center}
  \includegraphics[width=\textwidth]{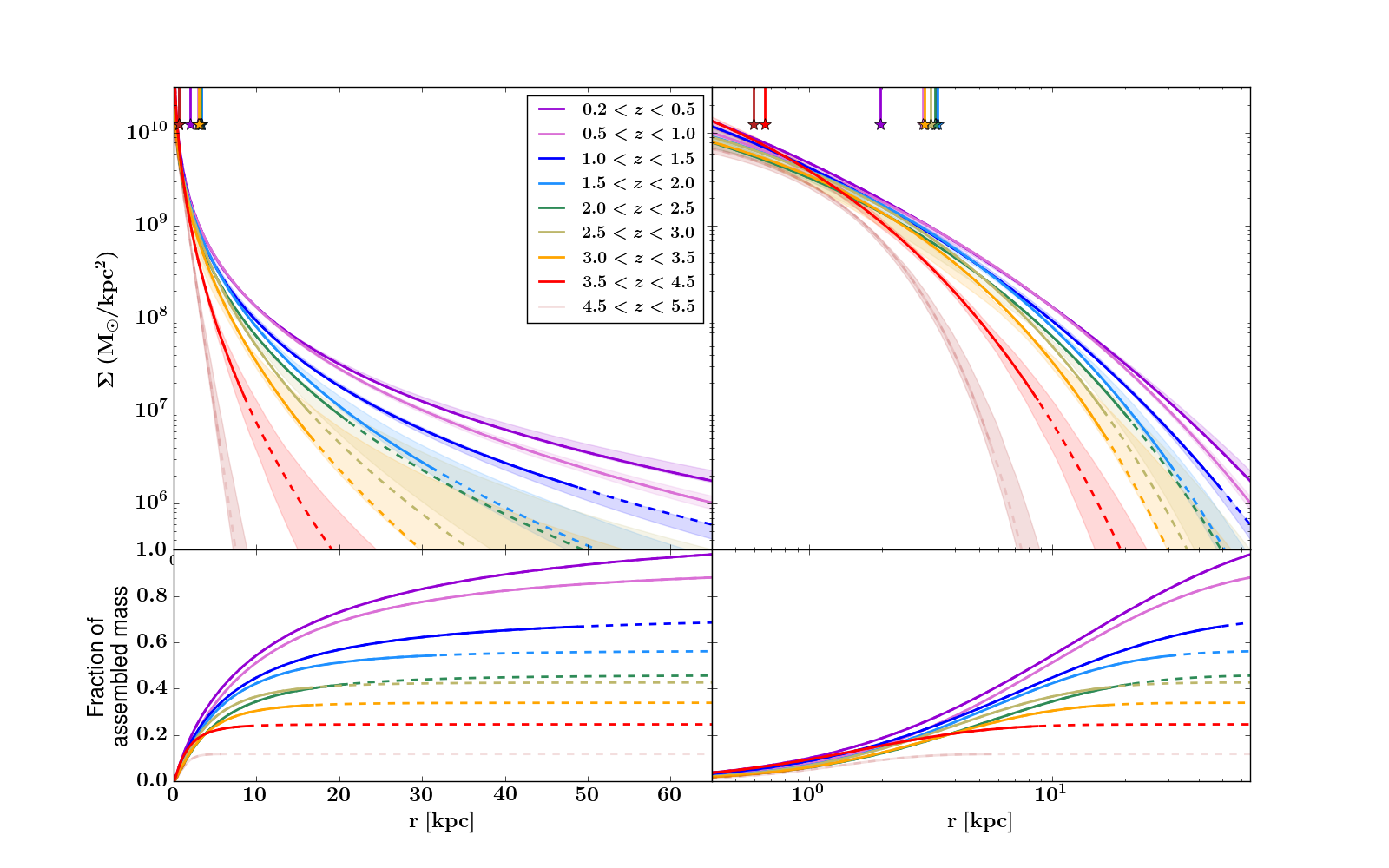}
  \caption{This figure is analogous to Fig.~\ref{fig:sb}, with the profiles derived from stacks of galaxies using a fixed cumulative number density selection. In this figure, we see that the increase in the surface mass density within $1-2~\mathrm{kpc}$ observed in Fig.~\ref{fig:sb} largely disappears, and the inner profiles do not show strong evolution with redshift.}
  \label{fig:sb_fixed}
  \end{center}
\end{figure*}

To properly investigate this, we repeated our analysis (as detailed in Sec.~\ref{sec:analysis}) for a fixed cumulative number density selected sample. In Fig.~\ref{fig:sb_fixed}, we plot the resultant surface mass density profiles. A comparison of the right panels of Fig.~\ref{fig:sb} and Fig.~\ref{fig:sb_fixed} shows the new mass selection significantly alters the observed surface mass density profiles in the central regions. In Fig.~\ref{fig:sb}, we see a difference of $\approx1~\mathrm{dex}$ between the lowest and highest redshift bin at $r<2~\mathrm{kpc}$. In contrast, the inner profiles in Fig.~\ref{fig:sb_fixed} lie approximately on top of each other with most mass evolution occurring in the outskirts.  

By choosing progenitors using the same methods as previous studies \citep[e.g.,][]{vandokkum2010, patel2013a, vandokkum2014}, we recover their trends, i.e., there is very little redshift evolution in the central stellar surface mass densities and that most mass evolution is occurring in the outskirts ($r>2~\mathrm{kpc}$). The effect of selection on the evolution of surface mass-density profiles is two-fold. First, a fixed cumulative number density selection yields higher mass progenitors, which will tend to be more spheroidal, and more centrally concentrated. Secondly, for an evolving cumulative number density selection, the mass evolves more steeply, with less-massive progenitors at high redshift. This will mean the normalization of surface-mass density profiles will also evolve more steeply which is reflected in the evolution of the central stellar surface mass density (as seen in Fig.~\ref{fig:sb}).

\end{document}